\documentclass[10pt]{book}

\usepackage{amsmath}
\usepackage{graphicx,epsfig}

\begin{document}
\pagestyle{empty}

\vspace*{-3cm}
\begin{flushleft}
\begin{minipage}[t]{11cm}
\hspace*{-2cm} Universit\'e Henri Poincar\'e\\
\hspace*{-2cm} Facult\'e des sciences et techniques\\
\hspace*{-2cm} UFR Sciences et Techniques de la Mati\`ere et des Proc\'ed\'es \\
\end{minipage}
\end{flushleft}

\vspace*{1cm}

\centerline{\large \sc Habilitation \`a Diriger des Recherches}

\vspace*{6cm}

\centerline{\Huge \bf Ising Quantum Chains}

\vspace*{1cm}

\centerline{\large \sc Dragi Karevski}
\centerline{Laboratoire de Physique des Mat\'eriaux, UMR CNRS 7556}
\vspace*{0.5cm}

\centerline{Soutenance publique le 14 d\'ecembre 2005}

\vspace*{3cm}
\noindent \hspace*{-1cm}  Membres du jury :\\

\noindent \hspace*{-1cm} Rapporteurs :\\
\noindent Pr. Peter Holdsworth, LPS, ENS Lyon \\
\noindent Pr. Yurij Holovatch, ICMP, Lviv,  Ukraine \\
\noindent Pr. Gunter Sch\"utz, IF, Forschungszentrum J\"ulich, Allemagne \\
\noindent \hspace*{-1cm} Examinateurs :\\
\noindent Pr. Malte Henkel, LPM, UHP Nancy, UMR CNRS 7556 \\ 
\noindent Pr. Wolfhard Janke, ITP, Universit\'e de Leipzig, Allemagne \\ 
\noindent Pr. Lo\"{\i}c Turban, LPM, UHP Nancy,  UMR CNRS 7556\\

\title{\sc Ising quantum chains}
\newpage

\begin{flushright}

 \begin{minipage}[t]{9cm}
{\small \it         Le travail est d\'esormais assur\'e d'avoir toute la bonne  
                       conscience de son c\^ot\'e : la propension \`a la joie se 
		       nomme d\'ej\`a ``besoin de repos"  et commence \`a se 
		       ressentir comme un sujet de honte. [...] Oui, il se pourrait 
		       bien qu'on en v\^{\i}nt \`a ne point c\'eder \`a un penchant 
		       pour la {\rm vita contemplativa} (c'est-\`a-dire pour aller se 
		       promener avec ses pens\'ees et ses amis) sans mauvaise 
                       conscience et m\'epris de soi-m\^eme.} \\ 
\begin{flushright}
{\small  Friedrich {\sc  Nietzsche}}\\
\small   Le Gai Savoir \\
~\\
\end{flushright}
\end{minipage}
\end{flushright}

\newpage

\tableofcontents

\chapter*{Introduction}
\addcontentsline{toc}{chapter}{Introduction}

Quantum spin chains are probably the simplest quantum mechanical systems showing a wide variety of
interesting properties \cite{lieb,schulz,haldane1,haldane2,nagaosa}, a main one being
the existence of quantum phase transitions, that is transitions at zero temperature
driven by large quantum fluctuations \cite{sachdev}. Singularities occuring strictly \cite{cardy}
at the zero-temperature
transition point can, however, produce a typical signature at (very) low temperatures.
On the experimental side, physicists are nowadays able to produce artificial samples
with behaviour that fits very well with the theoretical descriptions \cite{steiner,lemmens,bitko}.

The low-dimensionality of these systems allows the use of efficient analytical and numerical
tools such as the Bethe ansatz \cite{bethe,gaudin}, bosonization \cite{gogolin,vondelft},
or fermionization \cite{jordanwigner,lieb}, exact diagonalization of finite chains or the numerical
density matrix (DMRG) approach \cite{white,white1,peschelkaulke}.
We focus our attention in this review on free-fermionic quantum spin chains, which are spin models
that can be mapped on systems of non-interacting fermions \cite{lieb}. The free nature of the elementary
excitations allows for an exact diagonalization of the Hamiltonian. It is not here necessary to
argue on the usefulness of exact solutions for the implementation of the general comprehension
we have of such many-body systems. In particular, they can be used as a {\it garde fou}
when dealing with more complex systems, unsolved or unsolvable. Moreover, some of the
properties they show can still be present in truly interacting systems, that is, models
that cannot be or have not yet been mapped on free particles or solvable problems.

Most of the studies during the last decades, have been devoted to understand the influence
of inhomogeneities \cite{igpetu}, such as
aperiodic modulation of the couplings between spins \cite{luck} or the presence of 
quenched disorder \cite{disorder}, on the nature of the phase transition \cite{berchechatelain}. This has
culminated in the work of D. Fisher \cite{fisher} on the random transverse field Ising quantum chain.
The extremely broad distribution of energy scales near the critical point, for such random chains,
allows the use of a decimation-like renormalization group transformation \cite{madasgupta}.
Another aspect which is considered in this paper is the non-equilibrium behaviour of such fermionic spin chains. The focus will be on homogeneous systems, since they allow analytical
calculations, although expressions are given which are valid for general coupling distributions.

The paper is organised as follows: in the two next chapters, we present the general features
of the free fermionic spin models and the canonical diagonalization procedure first introduced
by Lieb {\it et al.} \cite{lieb}. A detailed discussion is given on the excitation spectrum and
the associated eigenvectors. It is also pointed out how one can extract the full phase diagram
of such spin chains from the knowledge of the surface magnetization. Results on dynamical correlation 
functions are reviewed. 
Chapter \ref{chap_aper} gives some rapid introduction on the studies done on the aperiodic Ising quantum chain. 
We have focused the attention on the anisotropic scaling and the weak universality found in such systems.
In chapter \ref{chap_rand}, we present results obtained by ourself and others within the random Ising quantum chain problem. After quickly reviewing  RG-results worked out on normal homogeneous distributions of disorder, we present the results obtained on the critical behaviour of random chains with either L\'evy type disorder or inhomogeneous disorder. The attention is concentrated on the surface critical behaviour since, due to the particularly simple expression of the surface magnetization, it is possible to obtain many exact results for that quantity. 
The following chapter deals with the non-equilibrium behaviour.
After solving the Heisenberg equations of  motion for the basic dynamical variables, we present
some aspects of the relaxation of the transverse magnetization. We show that systems
with either conserved or non-conserved dynamics present, however, some similarities in their relaxation
behaviour. This is illustrated with the $XX$-chain for conserved dynamics, and with the Ising chain
for non-conserved dynamics. Two-time functions are also considered and aging, that is, the dependence 
of the relaxation process on the age of the system, is also discussed.

\pagestyle{headings}

\chapter{Ising Quantum chain }   

\section{Free fermionic models}
The generic $XY$ Hamiltonian that we will consider
is \cite{katsura,niemeijer,pfeuty} \index{$XY$ Hamiltonian}
\begin{equation}
H=-\frac{1}{2}\sum_{n=1}^{L-1}\left[\frac{1+\kappa}{2}\sigma_{n}^{x}\sigma_{n+1}^{x}+
\frac{1-\kappa}{2}\sigma_{n}^{y}\sigma_{n+1}^{y}\right]
-\frac{1}{2}\sum_{n=1}^{L}h\sigma_{n}^{z}
\label{eq1}
\end{equation}
where the
$$
\sigma_{n}=1\otimes 1 \dots \otimes \sigma \otimes 1\dots \otimes 1
$$
are Pauli matrices at site $n$ and $\kappa$ is an anisotropy parameter with
limiting values $\kappa=1$
corresponding to the Ising \index{Ising quantum chain} case with a $Z_{2}$ symmetry
and $\kappa=0$ describing the  $XX$-model \index{$XX$ quantum chain} which has $U(1)$ 
symmetry. We will consider here only free boundary conditions.
The phase diagram and the critical behaviour of this model are known exactly since the 
work of Barouch and McCoy in 1971 \cite{barouch,mccoybarouch}
who generalized results obtained previously in the case of a vanishing transverse 
field \cite{lieb}, or at $\kappa=1$ \cite{pfeuty}.
It was first considered in the framework
of conformal invariance \cite{henkel} in Ref.~\cite{gehlen}.
In this section, to give a self-consistent presentation, we present in full details the
diagonalization procedure,  following, more or less closely,
the initial work of Lieb {\it et al.} \cite{lieb}.

The Hamiltonian can be mapped exactly on a free fermion model, consisting of an assembly 
of non interacting Fermi-Dirac oscillators. To proceed, let us first introduce the ladder
operators
$$
\sigma^{\pm}=\frac{1}{2}(\sigma^{x}\pm {\rm i}\sigma^{y}) \; .
$$
In the diagonal basis of the $\sigma^{z}$ component, they simply act as
$\sigma^{+}|\downarrow\rangle=|\uparrow\rangle$ and 
$\sigma^{-}|\uparrow\rangle=|\downarrow\rangle$.
They satisfy the anticommutation rules at same site
$$
\{\sigma^{+},\sigma^{-}\}=1
$$
and by construction they commute at different sites. They look like Fermi operators apart 
that they commute on different sites. True  fermionic operators are obtained
through a Jordan-Wigner transformation \cite{jordanwigner}.

Consider a basis vector with all spins down (in $z$-direction) and use the notations
$|1\rangle\equiv|\uparrow\rangle$ and of course $|0\rangle\equiv|\downarrow\rangle$, 
then this state will be the vacuum state destroyed by all the lowering operators 
$\sigma^{-}$:
$$
\sigma_{n}^{-}|00\dots 0\rangle =0\, \qquad \forall n \; .
$$
From this vacuum state, all the other states can be built up by applying raising 
operators $\sigma^{+}$. The situation looks very much the same than with fermions. 
But to have really fermions we need antisymmetry, {\it i. e.} anticommutation rules 
also at different sites. We introduce new operators
\begin{eqnarray}
c_{n}=A(n)\sigma_{n}^{-}\; ,\nonumber \\
c_{n}^{+}=\sigma_{n}^{+}A^{+}(n)\nonumber
\end{eqnarray}
where $A(n)$ is a unitary operator commuting with $\sigma^{\pm}_n$:
$$
[A(n),\sigma_{n}^{\pm}]=0\; .
$$
Then automatically we have
$$
\{c_{n}^{+},c_{n}\}=\{\sigma_{n}^{+},\sigma_{n}^{-}\}=1
$$
and in particular
\begin{equation}
\sigma_{n}^{z}=2\sigma_{n}^{+}\sigma_{n}^{-}-1=2c_{n}^{+}c_{n}-1\; .
\end{equation}
In order to fulfil the antisymmetry principle (sign change under particle exchange), 
the creation and annihilation operators must satisfy \cite{mattuck}
\begin{eqnarray}
c_{l}^{+}|n_{1},n_{2},...,n_{l},...\rangle= 
(-1)^{\Sigma_{l}}(1-n_{l})|n_{1},n_{2},...,n_{l}+1,...\rangle\; ,\nonumber \\
c_{l}|n_{1},n_{2},...,n_{l},...\rangle(-1)^{\Sigma_{l}}n_{l}|n_{1},n_{2},...,n_{l}-1,...\rangle\; ,\nonumber
\end{eqnarray}
where $\Sigma_{l}=\sum_{i=1}^{l-1}n_{i}$ is the number of particles on the left of the 
$l$ site. We should find now an operator representation of the sign factor 
$(-1)^{\Sigma_{l}}$. It is obvious that the choice
$$
A(l)=\prod_{i=1}^{l-1}(-\sigma^{z}_{i})
$$
will perfectly do the job \cite{mattuck}. This leads finally to the so called 
Jordan-Wigner transformations \cite{jordanwigner} \index{Jordan-Wigner transformation}
\begin{equation}
c_{n}=\prod_{i=1}^{n-1}(-\sigma^{z}_{i})\sigma^{-}_{n}=\prod_{i=1}^{n-1}
\exp\left({\rm i}\pi\sigma^{+}_{i}\sigma^{-}_{i}\right)\sigma^{-}_{n}
\end{equation}
and the adjoint relation.
Inverting these relations, we obtain expressions of the ladder operators and the
original Pauli matrices in terms of the fermionic
creation and annihilation operators $c^{+}$ and $c$.
Replacing this into our Hamiltonian we obtain the quadratic form
\begin{equation}
H=\sum_{n,m=1}^{L}c_{n}^{+}A_{nm}c_{m}+\frac{1}{2}(c_{n}^{+}B_{nm}c_{m}^{+}-
c_{n}B_{nm}c_{m})
\end{equation}
where the real matrices $A$ and $B$ are respectively symmetric and antisymmetric 
since the Hamiltonian is hermitian.
The quadratic nature of the Hamiltonian in terms of the Fermi operators insures the 
integrability of the model. As in the bosonic case, this form can be diagonalized
by a Bogoliubov transformation \cite{lieb,bogo}.

Let us mention here briefly that more general quantum spin-$1/2$ chains are not tractable
using this approach. For example, in the Heisenberg model \cite{bethe}
$$
H=J\sum_{n}  \vec{\sigma}_{n}\vec{\sigma}_{n+1}
$$
there are terms of the form
$$
\sigma^{z}_{n}\sigma^{z}_{n+1}\propto c^{+}_{n}c_{n}c^{+}_{n+1}c_{n+1}
$$
leading to interacting fermions. With longer range interactions, such as 
$\sigma_{n}^{x}\sigma_{n+p}^{x}$,
or magnetic fields in the $x$ or $y$ directions, another problem arises due to the
non-locality of the Jordan-Wigner transformations. For example a next nearest neighbour
interaction $\sigma_{n}^{x}\sigma_{n+2}^{x}$ generates quartic terms like
$$
c_{n}^{\mu}c_{n+1}^{+}c_{n+1}c_{n+2}^{\nu}
$$
where the $\mu$ and $\nu$ upper-scripts refer to either creation or annihilation 
operators. In the case of a magnetic field, let us say in the $x$ direction, we have 
additional terms proportional to the spin operators
$$
\sigma_{n}^{x}=\left(\prod_{i=1}^{n-1}(c_{i}^{+}+c_{i})(c_{i}^{+}-c_{i})\right)
(c_{n}^{+}+c_{n})
$$
which are clearly even worse. Nonlocal effects also appear when dealing with closed 
boundary conditions \cite{lieb,niemeijer}.

Nevertheless, the Hamiltonian (\ref{eq1}) is not the most general free-fermionic
expression. We can still add for example terms of the form
$\sigma_{n}^{x}\sigma_{n+1}^{y}-\sigma_{n}^{y}\sigma_{n+1}^{x}$, the so called
Dzyaloshinskii-Moriya interaction \cite{tsukada,derzhko} or play with the boundary 
conditions \cite{bilstein}.

\section{Canonical diagonalization}
We now come to the diagonalization of Hamiltonian (\ref{eq1}). For that purpose, we 
express the Jordan-Wigner transformation \index{Jordan-Wigner transformation}
in terms of Clifford \index{Clifford operators} operators \cite{hinrichsen} 
$\Gamma_{n}^{1}, \Gamma_{n}^{2}$:
\begin{eqnarray}
\Gamma_{n}^{1}&=&\left(\prod_{i=1}^{n-1}-\sigma_{i}^{z}\right)\sigma_{n}^{x}\; ,\nonumber\\
\Gamma_{n}^{2}&=&-\left(\prod_{i=1}^{n-1}-\sigma_{i}^{z}\right)\sigma_{n}^{y}
\end{eqnarray}
These operators are the $2L$-generators of a Clifford algebra since
\begin{equation}
\{\Gamma_{n}^{i},\Gamma_{k}^{j}\}=2\delta_{ij}\delta_{nk}\; , 
\quad \forall i,j=1,2\; ;  \forall n,k=1,...,L
\end{equation}
and are real operators, that is ${\Gamma_{n}^{i}}^{+}=\Gamma_{n}^{i}$. They can be viewed 
as non-properly normalised Majorana fermions.
The different terms in the original Hamiltonian are expressed as
\begin{equation}\begin{array}{l}
\sigma_{n}^{z}={\rm i}\Gamma_{n}^{1}\Gamma_{n}^{2}\\
\sigma_{n}^{x}\sigma_{n+1}^{x}=-{\rm i}\Gamma_{n}^{2}\Gamma_{n+1}^{1}\\
\sigma_{n}^{y}\sigma_{n+1}^{y}={\rm i}\Gamma_{n}^{1}\Gamma_{n+1}^{2}
\end{array}\; .
\end{equation}
Introducing the two-component spinor
$$
\Gamma_{n}=\left(\begin{array}{l}
\Gamma_{n}^{1}\\
\Gamma_{n}^{2}\end{array}
\right)
$$
one can write the Hamiltonian in the form
\begin{equation}
H=\frac{1}{4}\sum_{n=1}^{L-1}\Gamma_{n}^{\dagger}\big[ \sigma^{\bf y}+
{\rm i}\kappa  \sigma^{\bf x}\big]\Gamma_{n+1}+
\frac{1}{4}\sum_{n=1}^{L}\Gamma_{n}^{\dagger} h \sigma^{\bf y}\Gamma_{n}\; ,
\end{equation}
where $\Gamma_{n}^{\dagger}=(\Gamma_{n}^{1},\Gamma_{n}^{2})$ and
$$
{{ \sigma}^{\bf y}}=\left(\begin{array}{cc}
0&-{\rm i}\\
{\rm i}& 0\end{array}\right)\; \quad
{{ \sigma}^{\bf x}}=\left(\begin{array}{ll}
0&1\\
1& 0\end{array}\right)\;
$$
are the Pauli matrices, not to be confused with the initial spin operators. Introducing 
the $2L$-component operator
${\bf \Gamma}^{\dagger}=(\Gamma_{1}^{\dagger},\Gamma_{2}^{\dagger}...,
\Gamma_{L}^{\dagger})$, the Hamiltonian is given by
\begin{equation}
H=\frac{1}{4}{\bf \Gamma}^{\dagger}{\bf T}{\bf \Gamma}
\label{eq9}
\end{equation}
where $\bf T$ is a  $2L\times 2L$ hermitian matrix \cite{igturban,itks} given by \index{$\bf T$ Matrix}
\begin{equation}
{\bf T}=\left(\begin{array}{cccccc}
{\bf D} & {\bf F}& 0 & \dots  & & 0\\
{ \bf F}^{\dagger} &  {\bf D} &   { \bf F} & 0 &\dots &0\\
0& \ddots & \ddots & \ddots & 0 &  \\
& & & & & \\
& & & & & \\
0&\dots &0& { \bf F}^{\dagger} & {\bf D} &   { \bf F}\\
0& \dots & & 0&  { \bf F}^{\dagger} &  {\bf D}
\end{array}\right)
\label{eqT}
\end{equation}
with
\begin{equation}
{\bf D}=h \sigma^{\bf y}\; ,\quad {\bf F}=\frac{1}{2}\left(\sigma^{\bf y}+
{\rm i}\kappa \sigma^{\bf x}\right)\; .
\end{equation}

To diagonalize $H$, we introduce the unitary transformation matrix $\bf U$ build up on the
eigenvectors of the $\bf T$ matrix:\index{$\bf T$ Matrix}
$$
{\bf T}V_{q}=\epsilon_{q}V_{q}    \; , q=1,...,2L
$$
with the orthogonality and completeness relations
$$
\sum_{i=1}^{2L}V^{*}_{q}(i)  V_{q'}(i) =\delta_{qq'}\; , \quad
\sum_{q=1}^{2L}V^{*}_{q}(i)  V_{q}(i') =\delta_{ii'}\; .
$$
Inserting into (\ref{eq9}) the expression
${\bf T}={\bf U}{\bf \Lambda}{\bf U}^{\dagger}$
where ${\bf \Lambda}_{pq}=\epsilon_{q}\delta_{pq}$ is the diagonal matrix, one arrives at
$$
H=\frac{1}{4}{\bf \Gamma}^{\dagger}{\bf U}{\bf \Lambda}{\bf U}^{\dagger}{\bf \Gamma}
=\frac{1}{4}{\bf X}^{\dagger}{\bf \Lambda}{\bf X}
=\frac{1}{4}\sum_{q=1}^{2L}\epsilon_{q} x_{q}^{+}x_{q}
$$
with  the diagonal $2L$-component operator
$$
\bf X = U^{\dagger}\Gamma \; .
$$
We introduce now the following parametrisation, which will become clear later, of the 
eigenvectors $V_{q}$:
\begin{equation}
V_{q}=\frac{1}{\sqrt{2}}\left(\begin{array}{c}
\phi_{q}(1)\\
-{\rm i}\psi_{q}(1)\\
\phi_{q}(2)\\
-{\rm i}\psi_{q}(2)\\
\vdots\\
\phi_{q}(L)\\
-{\rm i}\psi_{q}(L)\end{array}\right)\; .
\end{equation}
Utilising this parametrisation, the operators $x_{q}$ and their adjoints are given by
$$
x_{q}=\frac{1}{\sqrt{2}}\sum_{i=1}^{L}\left[\phi_{q}^{*}(i)\Gamma_{i}^{1}+{\rm i}  
\psi_{q}^{*}(i)\Gamma_{i}^{2}\right]\; ,
$$
$$
x^{+}_{q}=\frac{1}{\sqrt{2}}\sum_{i=1}^{L}\left[\phi_{q}(i)\Gamma_{i}^{1}-{\rm i}  
\psi_{q}(i)\Gamma_{i}^{2}\right]\; .
$$
Now, if we consider the normalised operators
$$
d_{q}=\frac{1}{\sqrt{2}}x_{q}
$$
we can easily check that together with the adjoints $d^{+}_{q}$, they define Dirac Fermions,
that is they satisfy the anticommutation rules
\begin{equation}
\{d^{+}_{q},d_{q'}\}=\delta_{q,q'}\; ,\quad  \{d^{+}_{q},d^{+}_{q'}\}=0  
\; ,\quad  \{d_{q},d_{q'}\}=0  \; .
\end{equation}
For example, the first bracket is evaluated as
\begin{eqnarray}
\{d^{+}_{q},d_{q'}\}&=&\frac{1}{4}\sum_{i,j}\phi_{q}(i)\phi_{q'}^{*}(j)
\{\Gamma_{i}^{1},\Gamma_{j}^{1}\}
+\psi_{q}(i)\psi_{q'}^{*}(j)\{\Gamma_{i}^{2},\Gamma_{j}^{2}\}  \nonumber      \\
&-&{\rm i} \psi_{q}(i)\phi_{q'}^{*}(j)\{\Gamma_{i}^{2},\Gamma_{j}^{1}\}
+{\rm i} \phi_{q}(i)\psi_{q'}^{*}(j)\{\Gamma_{i}^{1},\Gamma_{j}^{2}\}    \nonumber
\end{eqnarray}
and using the anticommutation rules for the Clifford operators and the normalisation of 
the eigenvectors one is led to the above mentioned result.

Finally we have the free fermion Hamiltonian
\begin{equation}
H=\frac{1}{2}\sum_{q=1}^{2L}\epsilon_{q}d_{q}^{+}d_{q}    \; .
\end{equation}

We now take into account the particular structure of the $\bf T$ matrix. 
\index{$\bf T$ Matrix}
Due to the absence of the diagonal Pauli matrix $\sigma^{\bf z}$
in the expression of $\bf T$, the non-vanishing elements ${\bf T}_{ij}$ are those with $i+j$ 
odd, all even terms are vanishing. This means that by squaring the $\bf T$ matrix we can 
decouple the original $2L$-eigenproblem into two $L$-eigenproblems. The easiest way to see 
this is to rearrange the matrix $\bf T$ in the form
\begin{equation}
{\bf T}=\left(\begin{array}{cc}
0& {\bf C}\\
{\bf C}^{\dagger}& 0\end{array}
\right)
\end{equation}
where the $L\times L$ matrix $\bf C$ is given by
\begin{equation}
{\bf C}={\rm -i}\left(\begin{array}{ccccc}
h& J_{y} &  & &\\
J_{x} & h & J_{y} &  {\cal O}& \\
& J_{x} &\ddots & \ddots &  \\
 & {\cal O} & \ddots & \ddots & J_{y} \\
 & & &J_{x} &h
 \end{array}\right)
\end{equation}
with $J_{x}=(1+\kappa)/2$ and $J_{y}=(1-\kappa)/2$.
Squaring the $\bf T$ matrix \index{$\bf T$ Matrix}
gives\footnote{The supersymmetric structure appearing in the $T^2$ matrix has been used in 
Ref.~\cite{bercheigloi}.}
\begin{equation}
{\bf T}^{2}=\left(\begin{array}{cc}
{\bf C}{\bf C}^{\dagger}&0\\
0&{\bf C}^{\dagger}{\bf C}\end{array}
\right)\; .
\end{equation}
In this new basis the eigenvectors $V_{q}$ are simply given by
$$
V_{q}=\frac{1}{\sqrt{2}}\left(\begin{array}{c}
\phi_{q}(1) \\
\phi_{q}(2) \\
\vdots \\
\phi_{q}(L) \\
-{\rm i}\psi_{q}(1) \\
\vdots \\
-{\rm i}\psi_{q}(L)
\end{array}\right)     \; ,
$$
and together with ${\bf T}^{2}$ we finally obtain the decoupled eigenvalue equations
\begin{equation}
{\bf C}{\bf C}^{\dagger} \phi_{q}=\epsilon_{q}^{2}\phi_{q}
\end{equation}
and
\begin{equation}
{\bf C}^{\dagger}{\bf C} \psi_{q}=\epsilon_{q}^{2}\psi_{q} \; .
\end{equation}
Since the ${\bf C}{\bf C}^{\dagger}  $ and ${\bf C}^{\dagger}{\bf C}$ are real symmetric 
matrices, their eigenvectors can be chosen real and they satisfy completeness and 
orthogonality relations. This justifies the initial parametrisation of the vectors 
$V_{q}$ and one recovers the original formulation of Lieb, Schultz and Mattis \cite{lieb}.

Finally, one can notice another interesting property of the $\bf T$ matrix 
\index{$\bf T$ Matrix} which is related to the particle-hole symmetry \cite{igturban}.
Due to the off-diagonal structure of $\bf T$, we have
\begin{eqnarray}
-{\rm i}{\bf C}\psi_{q}=\epsilon_{q}\phi_{q}\; ,\nonumber \\
{\bf C}^{\dagger}\phi_{q}=-{\rm i}\epsilon_{q}\psi_{q}\label{cd1}
\end{eqnarray}
and we see that these equations are invariant under the simultaneous change 
$\epsilon_{q}\rightarrow -\epsilon_{q}$ and $\psi_{q}\rightarrow -\psi_{q}$.
So, to each eigenvalue $\epsilon_{q}\ge 0$ associated to the vector $V_{q}$ 
corresponds an eigenvalue $\epsilon_{q'}=-\epsilon_{q}$ associated to the vector
$$
V_{q'}=\frac{1}{\sqrt{2}}\left(\begin{array}{c}
\phi_{q}(1) \\
\phi_{q}(2) \\
\vdots \\
\phi_{q}(L) \\
{\rm i}\psi_{q}(1) \\
\vdots \\
{\rm i}\psi_{q}(L)
\end{array}\right)     \; .
$$
Let us classify the eigenvalues such as
$$
\epsilon_{q+L}=-\epsilon_{q} \quad  \forall q=1,...,L
$$
with $\epsilon_{q}\ge 0$ $\forall q=1,...,L$. Then the Hamiltonian can be written as
$$
H=\frac{1}{2}\sum_{q=1}^{L}\left(\epsilon_{q}d_{q}^{+}d_{q}
-\epsilon_{q}d_{q+L}^{+}d_{q+L}\right)
$$
where the operators with $q=1,...,L$ are associated to particles and the operators 
with $q=L+1,...,2L$ are associated with holes, that is negative energy particles. 
So that, by the usual substitution
\begin{eqnarray}
\eta_{q}^{+}=d_{q}^{+} \quad  \forall q=1,...,L\nonumber\\
\eta_{q}^{+}=d_{q+L} \quad   \forall q=1,...,L
\end{eqnarray}
we rewrite now the Hamiltonian in the form
\begin{equation}
H=\frac{1}{2}\sum_{q=1}^{L}\left(\epsilon_{q}\eta_{q}^{+}\eta_{q}
-\epsilon_{q}\eta_{q}\eta_{q}^{+}\right)
=\sum_{q=1}^{L}\epsilon_{q}\left[\eta_{q}^{+}\eta_{q}-\frac{1}{2}\right]\; .
\end{equation}

\section{Excitation spectrum and eigenvectors}
The problem now essentially resides in solving the two linear coupled equations 
$-{\rm i}{\bf C}\psi_{q}=\epsilon_{q}\phi_{q}$
and ${\rm i}{\bf C}^{\dagger}\phi_{q}=\epsilon_{q}\psi_{q}$.
We will present here the solutions of two particular cases, namely the $XY$-chain without 
field \cite{lieb,niemeijer} and the Ising quantum chain \index{Ising quantum chain}  
in a transverse field \cite{pfeuty}.
In the following we assume, without loss of generality, that the system size $L$ is even
number and $\kappa \ge 0$.

\subsection{$XY$-chain}
From ${\rm i}{\bf C}^{\dagger}\phi_{q}=\epsilon_{q}\psi_{q}$, we have the bulk equations
\begin{eqnarray}
\frac{1-\kappa}{2}\phi_{q}(2k-1)+\frac{1+\kappa}{2}\phi_{q}(2k+1)
&=&-\epsilon_{q}\psi_{q}(2k) \nonumber\\
\frac{1-\kappa}{2}\phi_{q}(2k)+\frac{1+\kappa}{2}\phi_{q}(2k+2)
&=&-\epsilon_{q}\psi_{q}(2k+1)
\label{Exc1}
\end{eqnarray}
Due to the parity coupling of these equations, we have two types of solutions:
\begin{equation}
\phi_{q}^{I}(2k)=\psi_{q}^{I}(2k-1)=0\; \quad \forall k
\end{equation}
and
\begin{equation}
\phi_{q}^{II}(2k-1)=\psi_{q}^{II}(2k)=0\; \quad \forall k
\end{equation}
In the first case, the bulk equations that remain to be solved are
$$
\frac{1-\kappa}{2}\phi^{I}_{q}(2k-1)+\frac{1+\kappa}{2}\phi^{I}_{q}(2k+1)=-\epsilon_{q}\psi^{I}_{q}(2k)\nonumber\\
$$
with the boundary conditions
\begin{equation}
\phi^{I}_{q}(L+1)=\psi^{I}_{q}(0)=0\; .
\end{equation}
Here we absorb the minus sign in equation (\ref{Exc1}) into the redefinition
$$
\widetilde{\psi}_{q}= -\psi_{q}\; .
$$
Using the ansatz $\phi^{I}_{q}(2k-1)=e^{iq(2k-1)}$ and 
$\widetilde{\psi}^{I}_{q}(2k)=e^{iq2k}e^{i\theta_{q}}$ to solve the bulk 
equations~(\ref{Exc1}), we obtain
\begin{equation}
\cos q + {\rm i} \kappa \sin q = \epsilon_{q} e^{i\theta_{q}}
\end{equation}
that is
\begin{equation}
\epsilon_{q}=\sqrt{\cos^{2}q+\kappa^{2}\sin^{2} q} \ge 0
\end{equation}
and the phase shift
\begin{equation}
\theta_{q}= \arctan (\kappa \tan q )\; ,
\end{equation}
with $0<\theta_{q}\le 2\pi$ to avoid ambiguity.
The eigenvectors associated to the positive excitations satisfying the boundary equations 
are then
\begin{equation}\begin{array}{l}
\phi_{q}^{I}(2k+1)=A_q\sin\left(q(2k+1)-\theta_{q}\right)\\
\widetilde{\psi}_{q}^{I}(2k)=A_q\sin(q2k)\; ,
\end{array}
\end{equation}
with
\begin{equation}
q(L+1)=n\pi+\theta_{q}
\end{equation}
or more explicitly
\begin{equation}
q=\frac{\pi}{L+1}\left(n+\frac{1}{\pi}\arctan \left(\kappa\tan q\right)\right)\; .
\label{Exc2}
\end{equation}
The normalisation constant $A_q$ is easy to evaluate and is actually dependent on $q$.
Using $\theta_{q}=q(L+1)+n\pi$, one can write $\phi^{I}_{q}$ in the form
$$
\phi_{q}^{I}(2k+1)=-A_q\delta_{q}\sin q(L-2k)\; ,
$$
where $\delta_{q}=(-1)^{n}$ is given by the sign of $\cos q(L+1)$.
The equation (\ref{Exc2}) has $L/2-1$ real solutions, that in the lowest order in $1/L$ 
are given by
\begin{equation}
q_{n}\simeq \frac{\pi}{L}\left(n-\nu_{n}\right)
\end{equation}
with
\begin{equation}
\nu_{n}=\frac{n}{L}-\frac{1}{\pi}
\arctan \left(\kappa\tan\left(\frac{n\pi}{L}\right)\right)\; , n=1,2,...,\frac{L}{2}-1\; .
\end{equation}
There is also a complex root of (\ref{Exc2})
\begin{equation}
q_{0}=\frac{\pi}{2}+{\rm i}v
\end{equation}
where $v$ is the solution of
\begin{equation}
\tanh v=\kappa \tanh [v(L+1)]\; .
\end{equation}
With the parametrisation $x=e^{-2v}$ and $\rho^{2}=\frac{1-\kappa}{1+\kappa}$, one is led 
to the equation
$$
x=\frac{\rho^{2}}{1-x^{L}(1-\rho^{2}x)}
$$
and the first nontrivial approximation leads to
\begin{equation}
x^{-1}\simeq \rho^{-2}-(1-\rho^{4})\rho^{-2(L-1)}\; .
\end{equation}
The excitation associated with this localised mode (see the form of $\phi_{q_{0}}$ and 
$\psi_{q_{0}}$ with $q_{0}=\pi/2+{\rm i}v$) is exponentially close to the ground state, 
that is
\begin{equation}
\epsilon_{q_{0}}\simeq (1+\rho^{2})\rho^{L}\; .
\end{equation}
From this observation, together with some weak assumptions, the complete phase diagram of 
the system can be obtained. We will discuss this point later.

The solutions of the second type satisfy the same bulk equations but the difference lies 
in the boundary conditions
$\phi_{q}^{II}(0)=\psi_{q}^{II}(L+1)=0$. The eigenvectors are given by
\begin{equation}
\begin{array}{l}
\phi_{q}^{II}(2k)=A_q\sin(q2k)\\
\widetilde{\psi}_{q}^{II}(2k+1)=A_q\sin(q(2k+1)+\theta_{q})=-A_q\delta_{q}\sin q(L-2k)\; ,
\end{array}
\end{equation}
with
\begin{equation}
q=\frac{\pi}{L+1}\left( n - \frac{1}{\pi}\arctan\left(\kappa\tan q\right)\right)\;
\end{equation}
which has $L/2$ real roots.
To the leading order, one gets
\begin{equation}
q_{n}=\frac{\pi}{L}\left(n-\nu_{n}\right)
\end{equation}
with
\begin{equation}
\nu_{n}=\frac{n}{L}+
\frac{1}{\pi}\arctan\left(\kappa\tan\left(\frac{n\pi}{L}\right)\right)\; , 
n=1,2,...,\frac{L}{2}
\end{equation}
which completes the solution of the $XY$-chain.

\subsection{Ising-chain}
The solution of the Ising  \index{Ising quantum chain} chain ($\kappa=1$) proceeds along 
the same lines \cite{pfeuty}. The bulk equations are
\begin{equation}
h\phi_{q}(k)+\phi_{q}(k+1)=\epsilon_{q}\widetilde{\psi}_{q}(k)
\end{equation}
with the boundary conditions
\begin{equation}
\phi_{q}(L+1)=\psi_{q}(0)=0\; .
\end{equation}
With the same ansatz as before, one arrives at
\begin{equation}
h+e^{iq}=\epsilon_{q}e^{i\theta_{q}}
\end{equation}
that is
\begin{equation}
\epsilon_{q}=\sqrt{(h+\cos q)^{2}+\sin^{2}q}
\end{equation}
and
\begin{equation}
\theta_{q}=\arctan\left(\frac{\sin q}{h+\cos q}\right)\; .
\end{equation}
Taking into account the boundary conditions, the solutions are readily expressed as
\begin{equation}
\begin{array}{l}
\phi_{q}(k)=A\sin(qk-\theta_{q})\\
\psi_{q}(k)=-A\sin (qk)
\end{array}
\end{equation}
where $q$ is a solution of the equation
\begin{equation}
q=\frac{\pi}{L+1}\left(n+\frac{1}{\pi}\arctan\left(\frac{\sin q}{h+\cos q}\right)\right)\; .
\label{Exc3}
\end{equation}
The eigenvectors can then also be written in the form
\begin{equation}
\begin{array}{l}
\phi_{q}(k)=-A(-1)^{n}\sin(q(L+1-k))\\
\psi_{q}(k)=-A\sin (qk)
\end{array}\; .
\end{equation}
In the thermodynamic limit, $L\rightarrow \infty$, for $h\ge 1$, the equation (\ref{Exc3}) 
gives rise to $L$ real roots. On the other hand, for $h<1$, there is also one complex root 
$q_{0}=\pi + i v$ associated to a localised mode
such that $v$ is solution of
\begin{equation}
\tanh(v(L+1))=-\frac{\sinh v}{h-\cosh v}\; .
\end{equation}
To the leading order, we have $v\simeq \ln h$. The eigenvectors associated to this localised 
mode are
\begin{equation}
\begin{array}{l}
\phi_{q_{0}}(k)=A(-1)^{k}\sinh(v(L+1-k))\\
\psi_{q_{0}}(k)=-A(-1)^{k}\sinh(vk)\; .
\end{array}
\end{equation}
Exactly at the critical value $h=1$, we have $\theta_{q}=q/2$, which gives a simple 
quantisation condition:
\begin{equation}
q=\frac{2n\pi}{2L+1}\; , n=1,2,...,L
\end{equation}
Changing $q$ into $\pi-q$, we have \cite{bercheturban}
\begin{equation}
\begin{array}{l}
\phi_{q}(k)=\frac{2}{\sqrt{2L+1}}(-1)^{k+1}\cos (q(k-1/2))\\
\psi_{q}(k)=\frac{2}{\sqrt{2L+1}}(-1)^{k}\sin(qk)\; ,
\end{array}
\end{equation}
and
\begin{equation}
\epsilon_{q}=2\left|\sin\frac{q}{2}\right|\; ,
\end{equation}
with $q=(2n+1)\pi/(2L+1)$ and $n=0,1,...,L-1$.

\chapter{Equilibrium behaviour}

\section{Critical behaviour}
From the knowledge of the eigenvectors $\phi$ and $\psi$, and the corresponding one-particle 
excitations, we can in principle calculate all the physical quantities, such as 
magnetization, energy density or correlation functions. However, they are, in general, 
complicated many-particles expectation values due to the non-local expression
of the spin operators in terms of fermions. We will come later to this aspect when 
considering the dynamics. Nevertheless, quantities that can be expressed locally in terms of
fermion are simple, such as correlations involving only $\sigma^{z}$ operators
or $\sigma_{1}^{x}$.

\subsection{Surface magnetization}
A very simple expression is obtained for the surface magnetization, that is the 
magnetization \index{Surface magnetization} on the $x$ (or $y$) direction of the first site. 
The behaviour of the first spin \cite{peschel,karevski1} gives general informations on the
phase diagram of the chain \cite{barouch,mccoybarouch,henkel}.
Since the expectation value of  the magnetization operator in the ground state vanishes, 
we have to find a bias. The usual way will be to apply a magnetic field in the desired 
direction, in order to break the ground state symmetry. Of course this procedure
has the disadvantage to break the quadratic structure of the Hamiltonian. Another route is 
to extract the magnetization behaviour from that of the correlation function. 
In this respect, the $x$($y$) component of the surface magnetization is obtained from the 
autocorrelation function
$G(\tau)=\langle \sigma^{x}_{1}(0)\sigma^{x}_{1}(\tau)\rangle$ in imaginary time $\tau$ 
where $\sigma^{x}_{1}(\tau)=e^{\tau H}\sigma^{x}_{1} e^{-\tau H}$.
Introducing the diagonal basis of $H$, we have
\begin{equation}
G(\tau)=|\langle\sigma|\sigma_{1}^{x}|0\rangle|^{2}e^{-\tau(E_{\sigma}-E_{0})}
+\sum_{i>1}|\langle i|\sigma_{1}^{x}|0\rangle |^{2}
e^{-\tau(E_{i}-E_{0})}
\end{equation}
where $|0\rangle$ is the ground state with energy $E_{0}$ and 
$|\sigma\rangle=\eta_{1}^{+}|0\rangle$
is the first excited state with one diagonal fermion whose energy is 
$E_{\sigma}=E_{0}+\epsilon_{1}$.
From the previous section we see that we have a vanishing excitation for $h<1$ in the 
thermodynamic limit $L\rightarrow \infty$ leading to a degenerate ground state. 
This implies that in the limit of large $\tau$, only the first term in the previous 
expression of the autocorrelation function contributes:
\begin{equation}
\lim_{\tau\rightarrow \infty} G(\tau)= [m_{s}^{x}]^{2}\; ,
\end{equation}
where $m_{s}^{x}=\langle\sigma|\sigma_{1}^{x}|0\rangle$.
Noticing that $\sigma^{x}_{1}=\Gamma_{1}^{1}=c^{+}_{1}+c_{1}$ and making use of the inverse 
expression,
$\Gamma_{n}^{1}=\sum_{q}\phi_{q}(n)(\eta_{q}^{+}+\eta_{q})$, one obtains
\begin{equation}
m_{s}^{x}=\langle\sigma|\sigma^{x}_{1}|0\rangle=\phi_{1}(1)\; .
\end{equation}
Similarly, one can obtain $m_{s}^{y}=\langle\sigma|\sigma^{y}_{1}|0\rangle=\psi_{1}(1)$.

Following Peschel \cite{peschel}, it is now possible to obtain a closed formula for the 
surface magnetization \cite{karevski1}. \index{Surface magnetization}
In the semi-infinite limit $L\rightarrow \infty$, for $h<1$, the first gap 
$E_{\sigma}-E_{0}=\epsilon_{1}$ vanishes due to
spontaneous symmetry breaking. In this case, equations (\ref{cd1}) simplify into
\begin{equation}
\begin{array}{l}
{\bf C}^{\dagger}\phi_{1}=0\\
{\bf C}\psi_{1}=0
\end{array}\; .
\label{eq58}
\end{equation}
Noticing that changing $\kappa$ into $-\kappa$, $m^{x}$ and $m^{y}$ are exchanged, in the 
following we will consider only the $x$-component.

To find the eigenvector $\phi_{1}$, we rewrite ${\bf C}^{\dagger}\phi_{1}=0$ in the 
iterative form
\begin{equation}
\left(\begin{array}{c}
\phi_{1}(n+1)\\
\phi_{1}(n)\end{array}\right)= {\bf K}_{n}\left(\begin{array}{c}
\phi_{1}(n)\\
\phi_{1}(n-1)\end{array}\right)
\end{equation}
where ${\bf K}_{n}$ is a $2\times 2$ matrix associated to the site $n$ whose expression for 
homogeneous coupling constants is
\begin{equation}
{\bf K}_{n}={\bf K}=-\left(\begin{array}{cc}
\frac{2h}{1+\kappa}& \frac{1-\kappa}{1+\kappa}\\
-1 & 0\end{array}\right)\; .
\end{equation}
By iterations, we obtain for the $(n+1)$th component of the eigenvector $\phi_{1}$ the 
expression
\begin{equation}
\phi_{1}(n+1)=(-1)^{n}\phi_{1}(1)\left({\bf K}^{n}\right)_{11}
\end{equation}
where the indices $11$ stand for the 1,1 component of the matrix ${\bf K}^{n}$. 
The normalisation of the eigenvector,
$\sum_{i}\phi_{1}^{2}(i)=1$, leads to the final expression \cite{karevski1}
\begin{equation}
m^{x}_{s}=\left[1+\sum_{n=1}^{\infty}\left|\left({\bf K}^{n}\right)_{11}\right|^{2}\right]^{-1/2}\; .
\label{cb1}
\end{equation}
First we note that the transition from a paramagnetic
to an ordered phase is characterised by the divergence of the sum entering (\ref{cb1}).
On the other hand, since for a one-dimensional quantum system with short-range interactions, 
the surface cannot order by itself, the surface transition is the signal of a transition in 
the bulk. It means that one can obtain some knowledge of the bulk by studying surface 
quantities. To do so, we first diagonalize the $\bf K$ matrix. The eigenvalues are
\begin{equation}
\lambda_{\pm}=\frac{1}{1+\kappa}\left[h\pm \sqrt{h^{2}+\kappa^{2}-1}\right]
\end{equation}
for $h^{2}+\kappa^{2}>1$ and complex conjugates otherwise
\begin{equation}
\lambda_{\pm}=\frac{1}{1+\kappa}\left[h\pm {\rm i} \sqrt{1-h^{2}-\kappa^{2}}\right]\rho \exp(\pm {\rm i}\vartheta)
\end{equation}
with $\rho=\sqrt{(1-\kappa)/(1+\kappa)}$ and $\vartheta=\arctan(\sqrt{1-h^{2}-\kappa^{2}}/h)$
and they become degenerate on the line $h^{2}+\kappa^{2}=1$. 
The leading eigenvalue gives the behaviour of $\phi_{1}^{2}(n)\sim |\lambda_{+}|^{2n}$
which, for an ordered
phase, implies that $|\lambda_{+}|<1$. The first mode is then localised near the surface. 
From this condition on $\lambda_{+}$,
we see that it corresponds to $h<h_{c}=1$. For $h>1$, the surface magnetization
\index{Surface magnetization} exactly vanishes and we infer that the bulk is not
ordered too. So that for any anisotropy $\kappa$, the critical line is at $h=1$ and in fact 
it belongs to the $2d$-Ising \cite{onsager} universality class.\footnote{To get an account on
the connection between $d$-dimensional quantum systems and $d+1$-dimensional classical
systems, on can refer to Kogut's celebrated review \cite{kogut}. The idea lies in the fact
that, in imaginary (Euclidean) time $\tau$, the evolution operator $e^{-\tau H}$, where $H$
is the Hamiltonian of the $d$-dimensional quantum system, can be interpreted as the transfer
matrix of a $d+1$ classical system.}
As it is shown hereafter, the special line $h^{2}+\kappa^{2}=1$, where the two eigenvalues 
collapse, separates an ordinary ferromagnetic phase ($h^{2}+\kappa^{2}>1$) from an 
oscillatory one ($h^{2}+\kappa^{2}<1$) \cite{barouch}. This line is known as the disorder line 
of the model.
The line at vanishing anisotropy, $\kappa=0$, is a continuous transition line (with 
diverging correlation length) called the anisotropic transition where the magnetization 
changes from $x$ to $y$ direction.

In the ordered phase, the decay of the eigenvector $\phi_{1}$ gives the correlation length 
of the system, which is related to the leading eigenvalue by
\begin{equation}
\phi^{2}_{1}(n)\sim |\lambda_{+}|^{2n}\sim \exp\left(-\frac{n}{\xi}\right)
\end{equation}
so that
\begin{equation}
\xi =\frac{1}{2|\ln|\lambda_{+}||}\; .
\end{equation}
By analysing $\lambda_{+}$, it is straightforward to see that the correlation length
exponent defined as $\xi\sim \delta^{-\nu}$, with $\delta\propto 1-h$ for the Ising transition
and $\delta\propto\kappa$ for the anisotropic one, is $\nu=1$.
Of course, a specific analysis of formula (\ref{cb1}) leads to the behaviour of the surface 
magnetization.\index{Surface magnetization}
At fixed anisotropy $\kappa$, close enough to the Ising transition line we have
\begin{equation}
m^{x}_{s}\sim (1-h)^{1/2}
\end{equation}
giving the surface critical exponent $\beta^{I}_{s}=1/2$.
In the oscillatory phase, a new length scale appears given by $\vartheta^{-1}$.
A straightforward calculation gives for the eigenvector $\phi_{1}$ the expression
\begin{equation}
\phi_{1}(n)=(-1)^{n}\rho^{n}\frac{1+\kappa}{h\tan \vartheta}\sin(n\vartheta) \phi_{1}(1)
\end{equation}
which leads to
\begin{equation}
m_{s}^{x}\sim \kappa^{1/2}
\end{equation}
close to the anisotropy line, so $\beta_{s}^{a}=1/2$ too.

We have seen here how from the study of surface properties \cite{karevski1} 
one can determine very simply (by the diagonalization of a $2\times 2$ matrix) 
the bulk phase diagram and also the exact correlation length \cite{barouch}. 
One may also notice that expression (\ref{cb1}) is suitable for finite
size analysis \cite{barber}, cut off at some size $L$. In fact for the Ising chain,
\index{Ising quantum chain}
to be precise, one can work with symmetry breaking boundary conditions \cite{igloirieger97}.  
That is, working
on a finite chain, we fix the spin at one end (which is equivalent to set $h_{L}=0$) and 
evaluate the magnetization on the other end. In this case, the surface magnetization 
\index{Surface magnetization} is exactly given by (\ref{cb1}) where the sum is truncated 
at $L-1$. The fixed spin at the end of the chain, let us say $\sigma_{L}^{x}=+$, leads to
an extra Zeeman term since we have now in the Hamiltonian the term 
$-\frac{J_{L-1}}{2}\sigma_{L-1}^{x}$, where the last coupling $J_{L-1}$ plays the role
of a magnetic field. The vanishing of $h_{L}$ induces a two-fold degeneracy of the 
Hamiltonian which is due to the exact vanishing of one excitation, say $\epsilon_{1}$. 
This degeneracy simply reflects the fact that $[\sigma_{L}^{x},H]=0$. On a mathematical 
ground, we can see this from the form of the $\bf T$ matrix \index{$\bf T$ Matrix}
which has a bloc-diagonal structure with a vanishing $2 \times 2$ last bloc. From the vacuum 
state associated to the diagonal fermions, $|0\rangle$, 
and its degenerate state $\eta_{1}^{+}|0\rangle$, we can form the two ground states
\begin{equation}
|\pm\rangle=\frac{1}{\sqrt{2}}\left(|0\rangle \pm \eta_{1}^{+}|0\rangle\right)
\end{equation}
associated respectively to $\sigma_{L}^{x}=+$ and $\sigma_{L}^{x}=-$. Since we have a 
boundary symmetry breaking field, we can directly calculate the surface magnetization from 
the expectation value of $\sigma_{1}^{x}$ in the associated ground state.
It gives $\langle +|\sigma_{1}^{x}|+\rangle=\phi_{1}(1)$, where $\phi_{1}(1)$ is exactly 
obtained for any finite size from ${\bf C}^{\dagger}\phi_{1}=0$. This finite-size expression
has been used extensively to study the surface properties of several inhomogeneous Ising
chains with quenched disorder \cite{igkari98,tukaig99,kaju,kalin}, where a suitable mapping
to a surviving random walk \cite{igloirieger97} problem permits to obtain exact results.

\subsection{Bulk magnetization}
As stated at the beginning of this section, quantities involving $\sigma^{x}$ or 
$\sigma^{y}$ operators are much more involved.
Nevertheless, thanks to Wick's theorem, they are computable in terms of 
Pfaffians \cite{caia,barouch} or
determinants whose size is linearly increasing with the site index. For example, 
if one wants to compute
$\langle \sigma|\sigma_{l}^{x}|0\rangle$, the
magnetization at site $l$, one has to evaluate the expectation value \cite{aper4}
\begin{equation}
m_{l}=\langle \sigma|\sigma_{l}^{x}|0\rangle\langle 0|\eta_{1}A_{1}B_{1}A_{2}B_{2}\dots A_{l-1}B_{l-1}A_{l}|0\rangle\; ,
\end{equation}
where we have defined $A_{i}=\Gamma_{i}^{1}$ and $B_{i}=-{\rm i}\Gamma_{i}^{2}$ 
in order to absorb unnecessary $\rm i$ factors.
These notations were initialy introduced by Lieb {\it et al} \cite{lieb}. 
Note that $B^{2}=-1$.
Since $A_{l}$ and $B_{l}$ are linear combinations of Fermi operators:
\begin{equation}
A_{l}=\sum_{q}\phi_{q}(l)\left(\eta^{+}_{q}+\eta_{q}\right)\; ,\quad
B_{l}=\sum_{q}\psi_{q}(l)\left(\eta^{+}_{q}-\eta_{q}\right)\; ,
\end{equation}
we can apply Wick's theorem \index{Wick Theorem} for fermions \cite{lieb}. The theorem states 
that we may expand the canonical (equilibrium) expectation value, with respect to a bilinear 
fermionic Hamiltonian, of a product of operators obeying anticommutation rules, in terms of 
contraction pairs. For example, if we have to evaluate the product 
$\langle C_{1}C_{2}C_{3}C_{4}\rangle$, we can expand it as
$$
\langle C_{1}C_{2}C_{3}C_{4}\rangle=\langle C_{1}C_{2}\rangle\langle C_{3}C_{4}\rangle
-\langle C_{1}C_{3}\rangle\langle C_{2}C_{4}\rangle
+\langle C_{1}C_{4}\rangle\langle C_{2}C_{3}\rangle\; .
$$
Due to the fermionic nature of the operators involved, a minus sign appears at each 
permutation. In our case, it is easy to see that the basic contractions 
\index{Pair contractions} $\langle 0| A_{i}A_{j}|0 \rangle$ and
$\langle 0|B_{i}B_{j}|0\rangle$
are vanishing for $i\neq j$. The only contributing terms are those products involving only
pairs of the type $\langle 0|\eta_{1}A|0\rangle$, $\langle 0|\eta_{1}B|0\rangle$ or
$\langle 0|BA| 0\rangle$. One may also remark that it is unnecessary to evaluate terms of 
the form $\langle 0|\eta_{1}B|0\rangle$ since in this case there is automatically
in the product a vanishing term $\langle 0|A_{i}A_{j}|0\rangle$ with $i\neq j$. The 
simplest non-vanishing product appearing in the Wick expansion is \index{Wick Theorem}
$$
\langle 0|\eta_{1}A_{1}|0\rangle 
\langle 0|B_{1}A_{2}| 0\rangle\dots \langle 0|B_{l-1}A_{l}| 0\rangle\; .
$$
The local magnetization $m_{l}$ is then given by the $l \times l$ 
determinant \cite{aper4,igloirieger97}
\begin{equation}
m_{l}=\left| \begin{array}{ccccc}
H_{1}&G_{11}&G_{12}& \dots&G_{1l-1}\\
H_{2}&G_{21}&G_{22}& \dots&G_{2l-1}\\
\vdots&\vdots&\vdots&      &\vdots\\
H_{l}&G_{l1}&G_{l2}& \dots&G_{ll-1}\end{array}
\right|
\end{equation}
with
\begin{equation}
H_{j}=\langle 0|\eta_{1}A_{j}|0\rangle=\phi_{1}(j)
\end{equation}
and
\begin{equation}
G_{jk}=\langle 0|B_{k}A_{j}| 0\rangle=-\sum_{q}\phi_{q}(j)\psi_{q}(k)\; .
\end{equation}
This expression for the local magnetization enables to compute, at least numerically, 
magnetization profiles \cite{tuig97,katuig} and to extract scaling behaviour.
For example far in the bulk of the
Ising chain  \index{Ising quantum chain} we have near the transition the power law behaviour 
$m_{b}\sim (1-h)^{1/8}$ for $h<1$ and zero otherwise \cite{pfeuty}. To conclude this section, 
one can also consider more complicated quantities such as two-point correlation 
functions \cite{lieb,pfeuty} in the same spirit.

\section{Time-dependent correlation functions}
Time-dependent correlations functions are of primary importance since experimentally
accessible dynamical quantities are, more or less simply, related to them. For general spin 
chains, the exact analysis of the long time behaviour of spin-spin correlations is a 
particularly difficult task. The generic time-dependent spin-spin correlation function is
$$
\langle \sigma_{i}^{\mu}(t)\sigma_{j}^{\nu}\rangle
$$
where $i,j$ are space indices, $\mu,\nu=x,y,z$ and where
the average $\langle\; .\; \rangle\equiv Tr\{\;.\;e^{-\beta H}\}/Tr\{e^{-\beta H}\}$ is the canonical
quantum expectation at temperature $T=1/\beta$. The time-dependent operator
$\sigma_{i}^{\mu}(t)=e^{{\rm i}Ht}\sigma_{i}^{\mu}e^{-{\rm i}Ht}$ is given by the usual 
Heisenberg representation. Most of the approximation schemes developed so 
far \cite{gagliano1,gagliano2} are not really relevant for such many-body
systems, at least at finite temperature.
One is ultimately forced to go on numerical analyses, basically by exact diagonalization 
of very short chains, although recently there has been a significant numerical progress 
using time-dependent DMRG (Density Matrix Renormalization Group) procedure \cite{white2}.

Nontrivial exact solutions for time-dependent correlations do exist for free fermionic
spin chains \cite{jacoby1,capelperk,katsura2,mccoybarouch,abraham,vaidya}. due to the 
non-interacting nature of the excitations. For such chains, not only bulk regimes were
investigated but also boundary effects \cite{cruz,stolze1,stolze2,pesch,pesch2}.
On one hand, the $z-z$ correlations are easily calculable due to their
local expression in terms of the Fermi operators. They are basically fermion density 
correlation function. On the other hand, the $x-x$ correlations are much more involved 
since in the Fermi representation one has to evaluate string operators. Nevertheless, 
as for the static correlators, one may use Wick's theorem \index{Wick Theorem} to
reduce them to the evaluation of a Pfaffian, or determinant, whose size is linearly 
increasing with $i+j$.

For the $N$-sites free boundary isotropic $XY$-chain in a transverse 
field,\index{$XY$ Hamiltonian}
$$
H=\frac{J}{4}\sum_{j=1}^{N-1}(\sigma_{j}^{x}\sigma_{j+1}^{x}+\sigma_{j}^{x}\sigma_{j+1}^{x})
-\frac{h}{2}\sum_{j=1}^{N}\sigma_{j}^{z}\; ,
$$
the basic contractions at inverse temperature $\beta$ \index{Pair contractions}
are given by \cite{cruz}:
\begin{equation}
\begin{array}{l}
\langle A_{j}(t)A_{l}\rangle =\frac{2}{N+1}\sum_{q}\sin qj \sin ql 
\left(\cos \varepsilon_{q} t
-{\rm i}\sin\varepsilon_{q}t\tanh \frac{\beta\varepsilon_{q}}{2}\right)\\
\langle A_{j}(t)B_{l}\rangle =\frac{2}{N+1}\sum_{q}\sin qj \sin ql 
\left({\rm i}\sin \varepsilon_{q} t
-\cos\varepsilon_{q}t\tanh \frac{\beta\varepsilon_{q}}{2}\right)
\end{array}
\end{equation}
and the symmetry relations
\begin{equation}
\begin{array}{l}
\langle B_{j}(t)B_{l}\rangle=-\langle A_{j}(t)A_{l}\rangle \\
\langle B_{l}(t)A_{j}\rangle=-\langle A_{j}(t)B_{l}\rangle
\end{array}\;
\end{equation}
where the excitation energies $\varepsilon_{q}=J\cos q -h$ with 
$q=n\pi/(N+1)\; , n=1,\dots, N$.
With the help of these expressions, one is able to evaluate the desired time-dependent 
correlations, at least numerically.

The $z-z$ correlator, namely $\langle \sigma_{j}^{z}(t)\sigma_{l}^{z}\rangle$, is given 
in the thermodynamic limit at infinite temperature $T=\infty$ by \cite{goncales,cruz}
\begin{equation}
\langle \sigma_{j}^{z}(t)\sigma_{l}^{z}\rangle\left[J_{j-l}(Jt)-(-1)^{l}J_{j+l}(Jt)\right]^{2}
\label{zzinfini}
\end{equation}
where $J_{n}$ is the Bessel function of the first kind. One has to notice
that this result is field independent. The bulk behaviour is obtained by putting 
$j,l\rightarrow \infty$ and keeping $l-j$ finite. One has
\begin{equation}
\langle \sigma_{j}^{z}(t)\sigma_{l}^{z}\rangle=J_{j-l}^{2}(Jt)
\stackrel{Jt\gg 1}{\sim}t^{-1}\; ,
\end{equation}
leading to a power law decay in time.
For large time, the boundary effects lead to
\begin{equation}
\langle \sigma_{j}^{z}(t)\sigma_{l}^{z}\rangle\simeq \frac{8}{\pi J^{3} t^{3}}
\left[\sin^{2}\left(Jt-\frac{\pi r}{2}\right)\right]
l^{2}(l+r)^{2}
\end{equation}
with $r=j-l$ the distance between the two sites. Therefore, the decay in time changes from 
$t^{-1}$ to $t^{-3}$ near the boundaries \cite{cruz}.

In the low-temperature limit ($T=0$), we have a closed expression for $h\ge J$ which is 
time and site independent \cite{cruz}:
\begin{equation}
\langle \sigma_{j}^{z}(t)\sigma_{l}^{z}\rangle=1\;
\end{equation}
revealing an ordered ground state. This can be related to the fact that the ground state 
corresponds to a completely filled energy-band, since all the energy excitations are negative.
As already stated before, the calculation of the $z-z$ correlation function is an easy task 
and we will not go on other models.

In order to calculate the $x-x$ time dependent correlation functions for such free-fermionic 
chains, one has to evaluate
\begin{equation}
\langle \sigma_{j}^{x}(t)\sigma_{l}^{x}\rangle=\langle A_{1}(t)B_{1}(t)\dots 
A_{j-1}(t)B_{j-1}(t)A_{j}(t)
A_{1}B_{1}\dots A_{l-1}B_{l-1}A_{l}\rangle\; .
\end{equation}
In the high temperature limit ($T=\infty$), the bulk correlation function of the isotropic 
$XY$ chain in a transverse field $h$ is given by the Gaussian 
behaviour \cite{sur,jacoby1,capelperk}:
\begin{equation}
\langle \sigma_{j}^{x}(t)\sigma_{l}^{x}\rangle=\delta_{jl}\cos ht \exp -\frac{J^{2}t^{2}}{4}
\end{equation}
where $\delta_{jl}$ is the Kronecker symbol. The boundary effects are hard to be taken into 
account due to the fact that the Pfaffian to be evaluated is a Toeplitz 
determinant \cite{wu,jacoby1,jacoby2} that can be treated only for large order \cite{jacoby1}.
Nevertheless for a vanishing transverse field, a conjecture was inferred from exact 
calculations for the boundary nearest sites $i=1,2,\dots,5$, claiming an asymptotic 
power-like decay \cite{stolze1}:
\begin{equation}
\langle \sigma_{j}^{x}(t)\sigma_{l}^{x}\rangle\sim\delta_{jl}t^{-3/2-(j-1)(j+1)}\; .
\end{equation}

At vanishing temperature and for $h\ge J$, the basic contractions can be evaluated in a 
closed form too. This leads to \cite{mccoyperk,mullershrock,cruz,abraham} 
\index{Pair contractions}
\begin{equation}
\langle \sigma_{j}^{x}(t)\sigma_{l}^{x}\rangle=\exp(-{\rm i}ht)\exp[{\rm i}(j-l)]J_{j-l}(Jt)
\end{equation}
for the bulk correlation function. The Bessel function gives rise to a $t^{-1/2}$ asymptotic
behaviour. So moving from $T=\infty$ to $T=0$ there is a dramatic change in the time decay 
behaviour \cite{cruz}. Finally, one is also able to take into account the free boundary 
effects. In this case, for large enough time, the behaviour changes from $t^{-1/2}$ to 
$t^{-3/2}$ \cite{cruz}.

At finite temperature the asymptotic decay of the time-dependent correlators is 
exponential \cite{its}:
\begin{equation}
\langle \sigma_{i}^{x}(t)\sigma_{i+n}^{x}\rangle\propto t^{2(\nu_{+}^{2}+\nu_{-}^{2})}
\exp f(n,t)
\end{equation}
where $f(n,t)$ is a negative monotonically decreasing function with increasing $T$. 
The preexponents $\nu_{\pm}$ are known functions \cite{its} of the field $h$, the temperature 
and the ratio $n/t$. When $T\rightarrow \infty$, the function $f(n,t)$ diverges 
logarithmically indicating the change of the decay shape from exponential to 
Gaussian \cite{stolze2}.

To end this section, one may mention that the decay laws for the $x-x$ time-dependent 
correlation functions of the anisotropic $XY$ chain are basically the same as for the 
isotropic case, namely power law at zero temperature and Gaussian at infinite 
temperature \cite{jacoby2}.

\chapter{\label{chap_aper} Aperiodic modulations}

\section{Definition and relevance criterion}
Since the discovery of quasicrystals in the middle of the eighties \cite{quasicristals}, extensive studies on Ising quantum chains with quasiperiodic or aperiodically modulated couplings have been done in the nineties \cite{aper1,aper2,aper3,aper4,itks}, see also \cite{aperreview} for a review. The interest in the field of critical phenomena in such aperiodic quantum chains lied in the fact that they offered intermediate situations between pure and random cases.  

Let us first define the aperiodic modulation itself. Most of the aperiodic sequences considered were generated via inflation rules by substitutions on a finite alphabet, such that 
$A\rightarrow S(A)$, $B\rightarrow S(B)$, ..., where $S(A)$($S(B)$) is a finite word replacing the letter $A$($B$). Starting from an initial letter and generating the substitution {\it ad eternam}, one obtains an infinite sequence of letters to whom a coupling sequence of the chain can be associated by the rules $A\rightarrow J_A$, $B \rightarrow J_B$, ...
If one considers a two letter sequence then the cumulated deviation $\Delta(L)$ from the average coupling $J$ at a length scale $L$ is characterized by a wandering exponent $\omega$ such that
\begin{equation}
\Delta(L)=\sum_{i=1}^L(J_i-J)\sim \delta L^\omega
\end{equation}
where $\delta$ is the strength of the aperiodic modulation and where the wandering exponent $\omega$ is obtained from the substitution matrix $M$ whose elements $(M)_{ij}$ give the number of letters $a_j$ contained in the word $S(a_i)$ \cite{queffelec,dumont}. The wandering exponent is given by
\begin{equation}
\omega = \frac{\ln|\Lambda_2|}{\ln \Lambda_1}
\end{equation}
where $\Lambda_1$ and $\Lambda_2$ are the largest and next largest eigenvalues of the subsitution matrix.  Near the critical point of the pure system, the aperiodicity introduces a shift of the critical coupling
$\delta t \sim \xi^{\omega-1}\sim t^{-\nu(\omega-1)}$ which has to be compared to the distance to criticality $t$ \cite{luck}. This leads to Luck's criterion
\begin{equation}
\frac{\delta t}{t}\sim t^{-\Phi} \qquad \Phi=1+\nu(\omega-1)
\end{equation}
and in the Ising universality class since $\nu=1$, one obtains 
\begin{equation}
\Phi=\omega \; .
\end{equation}
When $\omega < 0$, the perturbation is irrelevant and the system is in the Onsager universality class. On the other hand for $\omega>0$, the fluctuations are unbounded and the perturbation is relevant. For $\omega=0$, that is in the marginal case, one may expect continuously varying exponents.

\section{Strong anisotropy, weak universality}
It is not here the purpose to give an exhaustive view on what was done in the context of aperiodic modulation of the Ising quantum chain, but rather
to exemplify some (simple) aspects of it which have not really been noticed before (as far as my knowledge goes).

The Ising quantum chain in a transverse field is defined by the hamiltonian:
\begin{equation}
{\cal H}=-\frac{1}{2}\sum_{i=1}^{L-1}J_{i}\sigma_{i}^{x}\sigma_{i+1}^{x}
-\frac{1}{2}\sum_{i=1}^{L}h_{i}\sigma_{i}^{z}\; ,
\label{eq1}
\end{equation}
where the $\sigma$'s are the Pauli spin operators and 
${J_{i}},{h_{i}}$ are inhomogeneous couplings. In 
marginal aperiodic systems, for which $h_{i}=h$ and 
$\lambda_{i}=\lambda R^{f_{i}}$ with $\lambda_{i}=J_{i}/h_{i}$ and $f_{i}=0,1$ generating the aperiodicity, 
an anisotropic scaling was found. For such systems, the smallest excitations  scale at the 
bulk critical point as $\Lambda\sim L^{-z}$ with the size $L$ of the chain.
In the marginal aperiodic case, it was shown numerically that the anisotropy exponent $z$ is continuously 
varying with the control parameter $R$ \cite{ani,aper4} 
\begin{equation}
z(R)=x_{m_s}(R)+x_{m_s}(R^{-1})\ge 1\; ,
\label{eq2}
\end{equation}
where $x_{m_s}(R)$ is the magnetic exponent associated to 
$m_{s}=\langle \sigma^{x}_{1}\rangle$.\footnote{One may notice that 
relation~(\ref{eq2}) holds for the homogeneous system, $R=1$, with $x_{m_s}=1/2$ and $z=1$.}
The observed symmetry in the exchange $R\leftrightarrow 1/R$ in~(\ref{eq2}) was demonstrated in~\cite{aper3}
for aperiodic systems generated by inflation rules,
using a generalization of an exact renormalization group method introduced first
in~\cite{igturban} and applied to several aperiodic systems in~\cite{itks}.
We show here that this equation comes from a relation, valid for any distribution of couplings  
leading to anisotropic scaling, that rely the first gap $\Lambda_{1}$ to the surface magnetization. 

Using a Jordan-Wigner transformation~\cite{lieb}, the hamiltonian (\ref{eq1}) can be rewritten in a 
quadratic form in fermion operators. It is then diagonalized by a canonical transformation
and reads
\begin{equation}
{\cal H}=\sum_{q=1}^{L}\Lambda_{q}(\eta_{q}^{\dag}\eta_{q}-\frac{1}{2})\; ,
\label{eq3}
\end{equation}
where $\eta_{q}^{\dag}$ and $\eta_{q}$ are the fermionic creation and anihilation operators. 
The one fermion excitations $\Lambda_q$ satisfy the following set of equations:
\begin{equation}
\Lambda_{q}\Psi_{q}(i)=-h_{i}\Phi_{q}(i)-J_{i}\Phi_{q}(i+1)\\
\Lambda_{q}\Phi_{q}(i)=-J_{i-1}\Psi_{q}(i-1)-h_{i}\Psi_{q}(i)
\label{eq4}
\end{equation}
with the free boundary condition $J_{0}=J_{L}=0$. The vectors $\Phi$ and $\Psi$ are related
to the coefficients of the canonical transformation and enter into the expressions of
physical quantities. For example the surface magnetization, 
$m_{s}=\langle\sigma_{1}^{x}\rangle$, is simply given by the first component of $\Phi_{1}$
associated to the smallest excitation of the chain, $\Lambda_{1}$.

Let us consider now a distribution of the couplings which leads to anisotropic scaling
with a dynamical exponent $z>1$, as it is the case for bulk marginal aperiodic modulation of the
couplings~\cite{ani,aper4,aper3,igturban,itks}. Then the bottom spectrum of the critical hamiltonian scales as
$
\Lambda_{q}\sim L^{-z}
$
in a finite size system. According to~Ref.\cite{aper3,igturban,itks}, the asymptotic size dependence of $\Lambda_{1}(L)$
is given by the expressions
\begin{equation}
\Lambda_{1}(L)\simeq (-1)^{L}\frac{\Psi_{1}(1)}{\Phi_{1}(L)}\prod_{i=1}^{L-1}\lambda_{i}^{-1}
\left[1+\sum_{i=1}^{L-1}\prod_{j=1}^{i}\lambda_{L-j}^{-2}\right]^{-1}
\label{eq6}
\end{equation}
\begin{equation}
\Lambda_{1}(L)\simeq (-1)^{L}\frac{\Phi_{1}(L)}{\Psi_{1}(1)}\prod_{i=1}^{L-1}\lambda_{i}^{-1}
\left[1+\sum_{i=1}^{L-1}\prod_{j=1}^{i}\lambda_{j}^{-2}\right]^{-1}\; ,
\label{eq7}
\end{equation}
which are valid at the critical point and in the ordered phase. 
Noting in~(\ref{eq6}) that
\begin{equation}
\prod_{i=1}^{L-1}\lambda_{i}^{-1}\left[1+\sum_{i=1}^{L-1}\prod_{j=1}^{i}\lambda_{L-j}^{-2}\right]^{-1}
=\prod_{i=1}^{L-1}\lambda_{i}\left[1+\sum_{i=1}^{L-1}\prod_{j=1}^{i}\lambda_{j}^{2}\right]^{-1}\; ,
\label{eq8}
\end{equation}
equation~(\ref{eq6}) becomes
\begin{equation}
\Lambda_{1}(L)\simeq (-1)^{L}\frac{\Psi_{1}(1)}{\Phi_{1}(L)}
\prod_{i=1}^{L-1}\lambda_{i}\left[1+\sum_{i=1}^{L-1}\prod_{j=1}^{i}\lambda_{j}^{2}\right]^{-1}\; .
\label{eq9}
\end{equation}
Now multiplying both sides of equation~(\ref{eq9}) with (\ref{eq7}) leads to 
\begin{equation}
\Lambda_{1}(L)\simeq
\left[1+\sum_{i=1}^{L-1}\prod_{j=1}^{i}\lambda_{j}^{-2}\right]^{-1/2}
\left[1+\sum_{i=1}^{L-1}\prod_{j=1}^{i}\lambda_{j}^{2}\right]^{-1/2}
\; .
\label{eq10}
\end{equation}
which is symmetric under the exchange $\lambda\leftrightarrow 1/\lambda$.
One recognizes in this expression the surface magnetization~\cite{peschel}
$m_{s}(L,\{\lambda_{i}\})=\left[1+\sum_{i=1}^{L-1}\prod_{j=1}^{i}
\lambda_{j}^{-2}\right]^{-1/2}$ of the quantum chain, so that  one finally obtains
\begin{equation}
\Lambda_{1}(L)\simeq m_{s}(L,\{\lambda_{i}\}) m_{s}(L,\{\lambda^{-1}_{i}\})\; .
\label{eq11}
\end{equation}
This relation connects a bulk quantity, $\Lambda_{1}$, with surface quantities, namely
$m_{s}(L,\{\lambda_{i}\})$ and the surface magnetization of the dual chain, $m_{s}(L,\{\lambda^{-1}_{i}\})$.

Consider now a deterministic distribution of the chain couplings, $\{\lambda_{i}\}$,
with $h_{i}=h$
and $\lambda_{i}=J R^{f_{i}}/h$ with $f_{i}$ following some sequence of $0$ and $1$.
The critical coupling $\lambda_{c}$ follows from the relation 
$\lim_{L \rightarrow\infty}\prod_{k=1}^{L}\left(J_{k}/h_{k}\right)^{1/L}=1$~\cite{pfeuty} and gives
$\lambda_{c}=R^{-\rho_{\infty}}$ with $\rho_{\infty}$ the asymptotic density of modified couplings $\lambda R$.
The modulation of the couplings introduces a perturbation which can be either
relevant, marginal or irrelevant. For the Ising quantum chain 
the fluctuations around the average coupling 
$\bar{\lambda}$ at a length scale $L$
$
\Delta(L)=\sum_{k=1}^{L}(\lambda_{k}-\bar{\lambda})\sim L^{\omega}
$
governs the relevance of the perturbation~\cite{luck}. According to Luck's criterion, 
if $\omega<0$ the fluctuations are bounded and the system is in the Ising
universality class. On the other hand for $\omega>0$, the fluctuations are unbounded and one has to
distinguish two different situations. 
The evaluation of the surface magnetization is related to the sum
$
\sum_{j=1}^{L}\lambda^{-2j}R^{-2n_{j}}
$
where $n_{j}$ is the number of modified couplings, $\lambda R$, at size $j$. At the critical point, the sum can be rewritten asymptotically as
$
\sum_{j=1}^{L} R^{-2Bj^{\omega}}
$.
Now assume that the coefficient $B$ is positif (if not the roles of $R>1$ and $R<1$ are reversed in 
the following discussion). For $R>1$ the previous sum is absolutely convergent for $L\to\infty$
and leads to a finite surface magnetization with exponentially small corrections in a finite size system. 
On the other hand for $R<1$, the sum is diverging exponentially
and the surface magnetization is governed by the dominant term $\exp(-2B\ln R L^{\omega})$, so that
\begin{equation}
m_{s}(L,R)\sim \exp\left(-A(R)L^{\omega}\right)\; ,
\label{eq13}
\end{equation}
with $A(R)>0$.
So from equation (\ref{eq11}), in both case ($R>1$ or $R<1$)
the first gap $\Lambda_{1}$ will show an essential singularity leading to $z=\infty$:
\begin{equation}
\Lambda_{1}(R,L)\sim \exp\left(-\tilde{A}(R)L^{\omega}\right)\; ,
\label{eq14}
\end{equation}
with $\tilde{A}(R)=A(R)$ for $R<1$ and $\tilde{A}(R)=A(1/R)$ for $R>1$
since from (\ref{eq11}) $\Lambda_{1}(R)=\Lambda_{1}(1/R)$.

In the marginal case, corresponding to $\omega=0$, that is a logarithmic divergence of the fluctuations,
$n_{j}\simeq\rho_{\infty}j+C\ln j$ where $C$ is some constant,
it can be shown that
the surface magnetization scales at the critical point as
\begin{equation}
m_{s}(L,\{\lambda_{i}\}_{c})\sim L^{-x_{m_s}(R)}
\label{eq15}
\end{equation}
with an exponent $x_{m_s}(R)$ varying continuously with the control
parameter $R$. In fact the sum can be evaluated at the critical point using $n_{j}\simeq\rho_{\infty}j+C\ln j$.
One obtains 
$\sum_{j=1}^{L}\lambda_{c}^{-2j}R^{-2n_{j}}
\simeq\sum_{j=1}^{L}R^{-2C\ln j}
\sim\int^{L}{\rm d}x \, x^{-2C\ln R}\sim L^{1-2C\ln R}$. This expression is only valid in the weak perturbation
regime for $R\simeq 1$, that is in first order in $\ln R$. For a stronger regime one has to retain higher terms
in the $n_{j}$ expression. At this order, the surface magnetic exponent is $x_{m_s}(R)\simeq 1/2-C\ln R$. 
One can remark that for a sequence like the period-doubling one~\cite{ani,aper4}, $x_{m_s}(R)=x_{m_s}(1/R)$ which implies
$C=0$ (one can test this numerically) and then the former calculation gives 
$x_{m_s}(R)=1/2+{\cal O}(\ln^{2} R)$.
Finally, from~(\ref{eq11}) and~(\ref{eq15}) one obtains the relation
\begin{equation}
z(R)=x_{m_s}(R)+x_{m_s}(R^{-1})\; .
\label{eq16}
\end{equation}
The anisotropy exponent $z$ is then given by one surface magnetic exponent $x_{m_s}$ which is a function
of the perturbation strength. The symmetry $R\leftrightarrow1/R$ of $z$ is due to the self-duality of the Ising
quantum chain which implies for all bulk quantities the relation $Q(\{\lambda_{i}\})=Q(\{\lambda_{i}^{-1}\})$.

For a symmetric distribution of couplings with respect to the center of the chain, leading to
$x_{m_s}(R)=x_{m_s}(1/R)$ (see the period-doubling case~\cite{ani,aper4,aper3,igturban,itks}), one observes
a weak universality for the surface critical behaviour. 
The bottom of the spectrum scales anisotropically as 
$\Lambda\sim L^{-z}\sim\xi_{\perp}^{-z}\sim t^{z\nu}$ where $t$
measures the deviation from the critical point and
$\nu=1$ is the exponent of the longitudinal correlation length $\xi_{\perp}$. So that from~(\ref{eq11})
\begin{equation}
m_{s}(t)\sim \left(t^{z}\right)^{1/2}\; .
\label{eq17}
\end{equation}
From anisotropic scaling, one obtains for the critical dimension of the surface energy density $e_{s}$
the scaling relation $x_{e_s}=z+2x_{m_s}$. Using the symmetry of $x_{m_s}$ one has
\begin{equation}
e_{s}\sim \left(t^{z}\right)^{2}\; .
\label{eq18}
\end{equation}
We see that we recover the homogeneous surface exponents $x_{m_s}=1/2$, $x_{e_s}=2$ when
the deviation from the critical point is mesured with $t^{z}\sim \Lambda_{1}$.

The same weak universality seems to hold for the bulk quantities. In fact, it was shown in~Ref.\cite{aper4} that
the bulk energy density scales as $e\sim L^{-z}$ for marginal aperiodic modulation of the couplings. 
So again the pure energy density exponent $x_{e}=1$ is recovered.
Here we have investigated the behaviour of the mean
bulk magnetization $m_{b}=1/L\sum_{i} m(i)$ for the period-doubling sequence
from finite size scaling analysis. The magnetization is evaluated at the 
bulk critical point for sizes up to $L=1024$. 
\begin{figure}[ht]
\epsfxsize=8truecm
\begin{center}
\mbox{\epsfbox{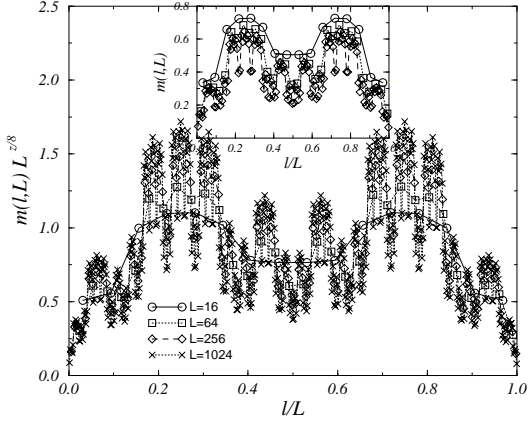}}
\end{center}
\caption{Rescaled magnetization profiles of the period-doubling chain with $r=5$. 
The insert gives the corresponding magnetization profile.}
\end{figure}
Numerically the profiles are well rescaled on the same curve with the exponent $x_{m_b}(R)=z(R)/8$ but
we notice that as the size increases the profiles are more and more decorated with a growing
fluctuation amplitude. 
This suggeres that the finite size behaviour of the mean critical magnetization is given by:
\begin{equation}
m_{b}\sim L^{-z/8}\sim \left(L^{-z}\right)^{1/8}\; .
\label{eq19}
\end{equation}
On the basis of the numerical datas, the magnetization profile is compatible with the form
\begin{equation}
m(l,L)=L^{-z/8}\left|\sin(\pi l/L)\right|^{x_{m_{s}}-x_{m_{b}}}[A+ G(l/L)]\; ,
\label{eq20}
\end{equation}
where $A$ is a constant and $G(x)$ is a kind of fractal Weierstrass function with zero mean value
whose fourier momentum are given by the period-doubling cascade.
The sinus term is very general for the profiles of the Ising quantum chains and is related to the geometry
of the system. This can be demonstrated
explicitely for the pure case \cite{tuig97} and was obtained numerically for random Ising systems \cite{igloi97a}.  The only difference here 
is that we have not only a pure constant in the front of it but in addition a fractal function of zero mean which controles
the local fluctuations, due to the aperiodic distribution, around some average environnement.

In conclusion the weak universality observed in this systems implies that the knowledge of 
the anisotropy exponent $z$ is sufficient to determine the critical behaviour of the system.

\chapter{\label{chap_rand}Disordered chains}

\section{Random transverse Ising chain}
\subsection{RG-equations}
Quenched disorder, i.e., time-independant disorder, has a strong influence on the nature of phase transitions and in particular on quantum phase transition. Many features of the randomness effects can be observed in one-dimensional systems, for which several exact results have been obtained and a general scenario (infinit randomness fixed point) has been developped~\cite{igmo}. In particular, since the work of Fisher~\cite{fisher}, the $1d$ random transverse Ising model has been the subject of much interest. The model is defined via the Hamiltonian:
\begin{equation}
{\cal H}=-\frac{1}{2}\sum_{i=1}^{L-1}J_{i}\sigma_{i}^{x}\sigma_{i+1}^{x}
-\frac{1}{2}\sum_{i=1}^{L}h_{i}\sigma_{i}^{z}\; ,
\label{eq4.1}
\end{equation}
where the exchange couplings, $J_i$ and transverse fields, $h_i$, are random variables.
In the case of homogeneous independently distributed couplings and fields, Fisher~\cite{fisher}
was able to obtained many new results about static properties of the model. The method used is based on a real space renormalization group (decimation like) method \cite{madasgupta}, which becomes exact at large scale, i.e., sufficiently close to the critical point (for a general extensive review of the method, one can refer to~\cite{igmo}). Later on, these results have been checked numerically and results have been obtained on the dynamical properties at the critical point and outside, in the so called Griffiths phase~\cite{igloi97a,rieger97b,igloirieger97,young96,mcke96,youn97,hegi98,fiyo98}.

Let us sketch the main properties of the random transverse Ising model:
For the inhomogeneous model, the phase transition is governed by the quantum control parameter 
\begin{equation}
\delta=[\ln h]_{av}-[\ln J]_{av}
\end{equation}
where the $[.]_{av}$ stands for an average over the quench disorder distribution.
The system is in a ferromagnetic phase in the region $\delta <0$, that is when the couplings are in average stronger than the transverse fields. Otherwise, it is in a paramagnetic phase.
The random critical point corresponds to $\delta=0$ and is governed by new critical exponents, that is a new universality class.

The renormalization procedure used by Fisher is a decimation-like method. It consists of the successive elimination of the strongest coupling (bond or field), which sets the energy scale of the renormalization (the energy cut-off), $\Omega=max\{J_i,h_i\}$. If the largest coupling is a bond, that is $\Omega=J_i$, then typically the randomly distributed neighboring fields $h_i$ and $h_{i+1}$ are much more smaller than $J_i$, i.e., $J_i\gg h_i,h_{i+1}$. Thus, one can treat perturbatively the two spins connected via the coupling $J_i$ and replace them by a single spin, with double momentum, in an effective field
$$
h_{i,i+1}=\frac{h_ih_{i+1}}{J_i}\; .
$$
If the strongest coupling is a field, $\Omega=h_k$, then the connecting bonds $J_{k-1}$ and $J_{k}$ to the neighboring spins are typically much smaller, i.e., $h_k\gg  J_{k-1},J_{k}$. From the magnetic point of view, this central spin is inactive, since due to the very strong transverse field, its susceptibility is very small and basically it gives no response to the total magnetization. Therefore, we can decimate out this spin and connect the remaining neighboring spins by an effective coupling
$$
J_{k-1,k}=\frac{J_{k-1}J_{k}}{h_k}\; .
$$
The renormalization consists of following the evolution of the relevant distributions 
during the lowering of the energy scale $\Omega$. The locality of the procedure and the fact that the couplings are random independently distributed, leads to integro-differential equations for the physical distributions that can be solved exactly. 

Let us examplify the discussion by considering the distributions of fields and bonds, respectively $P(h,\Omega)$ and $R(J,\Omega)$. Under the lowering of the energy scale from $\Omega$ to $\Omega-d\Omega$, a fraction  $d\Omega\left(P(\Omega,\Omega)+R(\Omega,\Omega)\right)$ of spins are eliminated. The total number of bonds with value $J$ at the new scale $\Omega-d\Omega$ is given by an equation of the type
$$
N(\Omega-d\Omega)R(J,\Omega-d\Omega)=N(\Omega)R(J,\Omega)+N(\Omega)F_{in}(J,\Omega)-
N(\Omega)F_{out}(J,\Omega)
$$
where, $N(\Omega)$ is the total number of spins at scale $\Omega$, and $F_{in,out}(J,\Omega)$ are the flux of incoming and outcoming bonds with value $J$ at the initial scale $\Omega$ during the decimation.
With $N(\Omega-d\Omega)/N(\Omega)=1-d\Omega\big(P(\Omega,\Omega)+R(\Omega,\Omega)\big)$, and accounting for the decimation process, one arrives at
\begin{eqnarray}
R(J,\Omega-d\Omega)\left[1-d\Omega\big(P(\Omega,\Omega)+R(\Omega,\Omega)\big)\right]=
\nonumber\\
R(J,\Omega)+d\Omega P(\Omega,\Omega)\int_0^{\Omega}dJ_1\int_0^{\Omega}dJ_2 \; R(J_1,\Omega)R(J_2,\Omega)\times \nonumber \\
\left\{\delta\left(J-\frac{J_1J_2}{\Omega}\right)-\delta(J-J_1)-\delta(J-J_2)\right\}
\end{eqnarray}
From duality, it is easy to obtain a similar RG-equation for the field distribution just by interchanging fields and bonds and $P$ and $R$. Expanding, $R(J,\Omega-d\Omega)$, one arrives at the integro-differential equations
\begin{eqnarray}
\frac{dR}{d\Omega}=R(J,\Omega)\big(P(\Omega,\Omega)-R(\Omega,\Omega)\big) \nonumber \\
-P(\Omega,\Omega)\int_j^{\Omega}dJ'\; R(J',\Omega)R(\frac{J}{J'}\Omega,\Omega) \frac{\Omega}{J'}\; ,
\end{eqnarray}
and a similar equation for the field distribution
\begin{eqnarray}
\frac{dP}{d\Omega}=P(h,\Omega)\big(R(\Omega,\Omega)-P(\Omega,\Omega)\big) \nonumber \\
-R(\Omega,\Omega)\int_h^{\Omega}dh'\; P(h',\Omega)P(\frac{h}{h'}\Omega,\Omega) \frac{\Omega}{h'}\; .
\end{eqnarray}
These are the basic equations, supplemented with the initial distributions, that have been solved by Fisher \cite{fisher}. The results obtained are asymptotically exact as the line of fixed points is approched
as $\Omega\rightarrow 0$.
We will not here continue into the details of the solution but rather we invite interested people to see ref. \cite{igmo} for a recent extensive review on strong disorder RG methods. 
To mention only few of them, the magnetization critical exponent $\beta$ is found to be $\beta=1-\tau$ where $\tau=\frac{\sqrt{5}-1}{2}$ is the golden mean and the dynamical exponent $z=1/(2|\Delta|)$ where $\Delta$ is the asymmetry parameter related to the relative strengths of the couplings and transverse fields. At the critical point, $\Delta=0$ and this leads to the log energy-scale $|\ln \Omega|\sim L^{\psi}$ for excitation energies on a finite system with size $L$.
The average spin-spin correlation function $G=[\langle \sigma^x_i\sigma^x_{i+r}\rangle]$ involves the average correlation length which diverges at the critical point as $\xi\sim \delta^{-\nu}$ with $\nu=2$ which differs from the typical value which is $\nu_{typ}=1$. These differences are due to the non self-averageness of the problem. 

\subsection{Surface behaviour}
The critical behaviour at the surface of the disordered chain can be obtained in a simple way, making a mapping to a random walk problem \cite{igloirieger97,kaju}. 
Indeed, fixing the end-spin $\sigma_L^x=+$, the surface magnetization is given by the formula \cite{peschel}
\begin{equation}
m_s=\left[1+\sum_{i=1}^{L-1}\prod_{j=1}^i \left(\frac{h_j}{J_j}\right)^2\right]^{-1/2}
\end{equation}
which is of Kesten random variable type. In the thermodynamical limit $L\rightarrow\infty$, from the analysis of the surface magnetization, one can infer that the system is in an ordered phase with finite average surface magnetization for 
\begin{equation}
\delta=[\ln h]_{av}-[\ln J]_{av} < 0
\end{equation}
where $[.]_{av}$ stands for an average over the disorder distribution. 
The system is in a paramagnetic phase for $\delta>0$. 

To be more concrete, let's put all the fields constant: $h_i=h_o\quad \forall i$ and let the couplings be defined as
\begin{equation}
J_i=\Lambda^{\theta_i}
\end{equation}
with $\Lambda > 0$ and
where the exponents $\theta_i$ are symmetric independent random variables. 
Due to the symmetry of the random variables $\theta$, the quantum control parameter is simply given by
\begin{equation}
\delta=\ln h_o
\end{equation}
so that the distance to the transition is controled only by the field value. The parameter $\Lambda$ controls the strength of the disorder.
For example, the binary distribution 
$$
\pi(\theta)=\frac{1}{2}\delta(\theta+1)+\frac{1}{2}\delta(\theta-1)
$$
corresponds to couplings either $J=\Lambda$ or $J=1/\Lambda$.
In the next subsection we will discuss the case where the distribution $\pi(\theta)$ is very broad with power law tails corresponding to so called L\'evy flights or Riemann walk. But at the present let us consider only normal distributions (satisfying central limit theorems).

The average surface magnetization, 
$[m_s]_{\rm av}(L)$, can be written as\cite{peschel}
\begin{equation}
[m_s]_{\rm av}(L)=\lim_{N\rightarrow\infty}\frac{1}{N}\sum_{n=1}^{N}
\left(\sum_{j=0}^{L}\exp\left[-2\ln\! \Lambda\; \{S^{(n)}_{j}-\delta_{w}j\}\right]\right)^{-1/2}
\label{e1.4}
\end{equation}
where the index $n$ referes the different disorder realizations and
with $\delta_{w}=\delta/\ln \Lambda$ and where the random sequence 
$S^{(n)}_{j}=\theta^{(n)}_{1}+\theta^{(n)}_{2}+\dots+\theta^{(n)}_{j}$ 
and $S^{(n)}_{0}=0$ by convention. Since the random variables $\theta_{k}$ are independent identically
distributed variables we can interpret $S_{j}$ as the displacement of a one dimensional random walker 
starting at the origin $S_{0}=0$ and arriving at a distance $S_{j}$ after $j$ steps. At each step the 
walker perform a jump of length $\theta$ distributed according to the symmetric distribution.
In the strong disorder limit, $\Lambda\gg 1$, the leading contribution to the average magnetization
comes only from those walks whose mean velocity $S_{j}/j$ is larger than the drift velocity $\delta_{w}$
up to time $L$, since then 
$1+\sum_{j=1}^{L}\exp(-2\ln\!\Lambda\; \{S_{j}-\delta_{w}j\})=O(1)$ and the corresponding contribution to the average 
magnetization is of order one. On the other hand when the walker mean velocity becomes smaller than $\delta_{w}$,
it gives an exponential contribution to the sum $\sum_{j=1}^{L}\exp(-2\ln\!\Lambda\; \{S_{j}-\delta_{w}j\})$
which is then dominated by the maximum negative displacement $S_{jmax}-\delta_{w}jmax$, so that the contribution to 
$[m_s]_{\rm av}$ is exponentially small, of order $O\left({\rm e}^{-\ln\!\Lambda\;\{|S_{jmax}-\delta_{w}jmax\}|}\right)$.
It means that the average surface magnetization $[m_s]_{\rm av}(L)$ is proportional to the surviving probability after $L$ steps of a walker which is absorbed if it crosses the $\delta_{w}j$ line. 
Stated differently, it means that the typical walks will give an exponentially small contribution to the average magnetization which is completely governed by rare events, that is non crossing walks.
The problem of the surface magnetization is then mapped onto a random walk problem with an absorbing boundary. 

In order to obtain the survival probability, 
we use Sparre Andersen formula \cite{andersen,andersen2}
\begin{equation}
F(z,\delta)=\exp \left(P(z,\delta)\right)
\label{e1.5}
\end{equation}
where $F(z,\delta)=\sum_{n\ge 0}F(n,\delta) z^{n}$ is the generating function of the survival probabilities $F(n,\delta)$
of the walker after $n$ steps such that the displacement of the walker $S_{j}>j\delta$ for all $j\le n$ and
$P(z,\delta)=\sum_{n\ge 1}(P(n,\delta)/n) z^{n}$ where  $P(n,\delta)$ is the probability that at step $n$, $S_{n}>n\delta$.
At the critical point, $\delta=0$, since the distribution is symmetric we have $P(n,0)\simeq1/2$ for $n\gg 1$
so that from Eq.(\ref{e1.5}) we have $F(n,0)\sim n^{-1/2}$ which implies \begin{equation}
[m_s]_{\rm av}(L)\sim L^{-1/2}\; ,
\end{equation}
that is $x_{m}^{s}=1/2$.

One is also interested in the typical behaviour at the critical point of the surface magnetization,
$m_{s}^{typ}(L)=\exp([\ln m_s(L)]_{\rm av})$. In the strong disorder regime, the leading contribution
to the typical magnetization comes from those walks who visit the negative axis, that is, walks not contributing
to the average magnetization. For such walks, $\ln m_{s}\simeq -\ln\!\Lambda\; |S_{jmax}|$, so that
\begin{equation}
m_{s}^{typ}(L)=\exp\left(-\ln\!\Lambda\; [|S_{jmax}|]_{\rm av}\right)
\label{e1.6}
\end{equation}
with $[|S_{jmax}|]_{\rm av}$ the average maximum displacement of the walker on the negative axis after $L$ steps.
This average is actually given by the absolute mean displacement which scales at large $L$ as $L^{1/2}$ and which measures the fluctuations of the walk. So that the typical magnetization is 
\begin{equation}
m_{s}^{typ}(L)\sim\exp(-{\rm const.} L^{1/2})\; ,
\end{equation}
which means that the appropriate scaling variable is $(\ln m_s)/L^{1/2}$ and one can show that the distribution of $m_s$ is logarithmically broad. 
We close here the discussion of the surface critical properties of the (normal) random Ising quantum chain. One can see ref.\cite{monthus} for exact results on the distribution functions of the surface magnetization. 

\subsection{Levy flights}
In the present subsection, we study the influence of very broad disorder distributions on the critical properties of the random Ising quantum chain \cite{kalin}. For that, we use a distribution $\pi(\theta)$ such that for large arguments it decreases
with a power law tails
\begin{equation}
\pi(\theta) \sim |\theta|^{-1-\alpha}\; , \quad |\theta| \gg 1
\end{equation}
where $\alpha$ is the so called L\'evy index. We will consider here the region $\alpha >1$ where the $k$-th moment of the distribution exists for $k<\alpha$.  
More precisely, we consider the distribution 
$\pi(\theta)=p {\alpha}  (1+\theta)^{-1-\alpha}$ for $\theta>0$ and
$\pi(\theta)=q {\alpha}  (1+|\theta|)^{-1-\alpha}$ for $\theta<0$, $p+q=1$. In ref.\cite{kalin} we have also considered the discretized version (Riemann walk) of the above distribution. Using the continuous distribution defined above, the quantum control parameter is given by $\delta=\ln h_{o}-(p-q)\ln \Lambda /(\alpha-1)$. Taking $h_o=1$, leads to $\delta=-(p-q)\ln \Lambda /(\alpha-1)$, which means that the asymmetry $p-q$ drives the system outside the critical point $\delta=0$. 

In the strong disorder limit $\Lambda\rightarrow \infty$, we have shown in the previous section that the average surface magnetization is given by the surviving probability of a random walk distributed according to $\pi(\theta)$ with an absorbing boundary moving with a constant velocity $\delta/\ln\Lambda$. 
So considering the following sum
\begin{equation}
S_n=\sum_{j=1}^n \theta_j\;,
\end{equation}
in the large $n$ limit, it exists a limit distribution $\tilde{p}(u)\rm{d}u$, in term of the variable, $u=S_n/l_n-c_n$ with 
the normalization
\begin{equation}
l_n=n^{1/\alpha}
\label{transv}
\end{equation}
giving the transverse fluctuation of the walk, if we interpret $n=t$ as the (discrete) time and $S_{n=t}$ as
the position of the walker in the transverse direction. The second normalization is given by
\begin{equation}
c_n=-n^{1-1/\alpha} \delta_w\;,
\label{longi}
\end{equation}
where with $\delta_w=-\langle \theta \rangle=\delta/\ln\Lambda$ we define the bias of the walk. 
For a small asymmetry $\delta_w$
one gets from the combination in Eq.(\ref{longi}) the scaling relation between time and bias as:
\begin{equation}
t \sim |\delta_w|^{-\nu(\alpha)},\quad \nu(\alpha)=\frac{\alpha }{\alpha-1}\;.
\label{nuwalk}
\end{equation}
We use here the notation $\nu$ for the exponent since remember that the final time $t$ plays the role of the chain size $L$.
For a symmetric distribution, when $\pi(\theta)$ is an even function, thus $p=q$ and $\delta_w=0$, we
have for the limit distribution:
\begin{equation}
\tilde{p}(u)=L_{\alpha,0}(u)=\frac{1 }{2 \pi} \int_{-\infty}^{\infty} e^{iku-|k|^{\alpha}} {\rm d}k\;,
\end{equation}
which has an expansion around $u=0$
\begin{equation}
L_{\alpha,0}(u)=\frac{1}{\pi \alpha} \sum_{k=0}^{\infty} (-1)^k \frac{u^{2k}}{(2k)!} \Gamma\left(\frac{2k+1}{ \alpha}
\right)\;
\label{smallu}
\end{equation}
and for large $u$ it is asymptotically given by
\begin{equation}
L_{\alpha,0}(u)=\frac{1 }{\pi} u^{-(1+\alpha)} \Gamma(1+\alpha) \sin(\pi \alpha/2)\;,
\label{largeu}
\end{equation}
where $\Gamma(x)$ denotes the gamma function.

Consider next the surviving probability, $P_{surv}(t,\delta_w)$, which is given by the fraction of
those walks, which have not crossed the starting position until $t=n$, thus $S_i>0$ for $i=1,2,\dots,n$.
For a biased walk, with $0<|\delta_w|\ll 1$, the asymptotic behaviour of $P_{surv}(n,\delta_w)$ is equivalent
to that of a symmetric walk ($\delta_w=0$) but with a moving adsorbing boundary site, which has a constant
velocity of $v=\delta_w$. For this event, with $S_i>vi$ for $i=1,2,\dots,n$, the surviving probability
is denoted by $F(n,v)$, whereas the probability for $S_n>vn$, irrespectively from the previous steps, is
denoted by $P(n,v)$ and the latter is given by:
\begin{equation}
P(n,v)=\int_{nv}^{\infty} p(S,n) {\rm d} S\;.
\label{pnv}
\end{equation}
Between the generating functions:
\begin{eqnarray}
F(z,v)&=&\sum_{n \ge 0} F(n,v) z^n\nonumber\\
P(z,v)&=&\sum_{n \ge 1} \frac{P(n,v)}{ n}z^n\;
\end{eqnarray}
there is a useful relation due to Sparre Andersen \cite{andersen,andersen2}:
\begin{equation}
F(z,v)=\exp\left[P(z,v)\right]\;,
\end{equation}
which has been used recently in Ref.\cite{bauer}.

In the zero velocity case, $v=0$, which is equivalent to the symmetric walk with $\delta_w=0$, we
have $P(n,0)$=1/2. Consequently $P(z,0)=-\frac{1 }{ 2} \ln(1-z)$ and $F(z,0)=(1-z)^{-1/2}$, from which
one obtains for the final asymptotic result:
\begin{equation}
P_{surv}(t,0)=\left. F(n,0)\right|_{n=t} \sim t^{-\vartheta},\quad\vartheta=1/2\;.
\label{symexp}
\end{equation}
Note that the persistence exponent, $\vartheta=1/2$, is independent of the form of a symmetric
probability distribution, $\pi(\theta)$, thus it does not depend on the L\'evy index, $\alpha$.
Thus, from the correspondence between the random walk problem and the random quantum chain, we obtain
\begin{equation}
[m_s(0,L)]_{av}\sim L^{-1/2}\; .
\end{equation}
Thus the anomalous dimension $x_m^s=1/2$ of the average surface magnetization does not depend on the L\'evy index $\alpha$ and its value is the same as for the normal distribution ($\alpha>2$).

For $v=\delta_w>0$ (paramagnetic case), i.e. when the allowed region of the particle shrinks in time the correction to
$P(n,0)=1/2$ has the functional form, $P(n,v)=1/2-g(\tilde{c})$, with $\tilde{c}=v n^{1-1/{\alpha}}$.
Evaluating Eq.(\ref{pnv}) with Eq.(\ref{smallu}) one gets in leading order,
$P(n,v)=\frac{1}{ 2} - \tilde{c} A(\alpha) +O(\tilde{c}^3)$,
with $A(\alpha)=\Gamma(1+1/\alpha)/\pi$. Then $P(z,v)-P(z,0)\simeq A(\alpha) v \sum_{n \ge 1} z^n n^{-1/\alpha}$
is singular around $z \to 1^-$ as $\sim (1-z)^{-(1-1/\alpha)}$, consequently
\begin{equation}
F(z,v)\simeq (1-z)^{-1/2} \exp\left[-A(\alpha) v (1-z)^{-(1-1/\alpha)} \right]\;,
\end{equation}
in leading order and close to $z=1^{-}$. Here the second factor gives the more singular contribution
to the surviving probability, which is in an exponential form:
\begin{eqnarray}
P_{surv}(t,\delta_w) &\sim &\left.F(n,v)\right|_{v=\delta_w,n=t} \nonumber\\
&\sim& t^{-1/2} \exp\left[ -{\rm const} \delta_w
t^{1-1/\alpha} \right]\;.
\label{psurv+}
\end{eqnarray}
Thus the average surface magnetization 
$\left[m_s(\delta,L)\right]_{\rm av} \sim \left. P_{surv}(t,\delta_w)\right|_{t=L}$
has an exponentially decreasing behaviour as a function of the scaling variable $\delta L^{1-1/\alpha}$.\footnote{In the Cauchy distribution case, corresponding to the limit $\alpha=1$, one can calculate exactly the surviving probability \cite{bauer}, and one finds $ F(t,\delta_w)\sim t^{-1/2-1/\pi \arctan \delta_w}$, which leads close to the critical point to $F(t,\delta_w)\sim t^{-1/2}\exp(-const.\delta_w\ln t)$ which has to be compared to the general $\alpha$ case.
}
Consequently the characteristic length-scale in the problem, the average correlation length, $\xi$,
and the quantum control parameter, $\delta$, close to the critical point are related as $\xi\sim \delta^{-\nu(\alpha)}$
with
\begin{equation}
\nu(\alpha)=\frac{\alpha}{\alpha -1}\;.
\label{nualpha}
\end{equation}
Note that $\nu(\alpha)$ is divergent as $\alpha \to 1^+$, which is a consequence of the fact that the
first moment of the L\'evy distribution is also divergent in that limit. In the other limiting case,
$\alpha \to 2^-$, we recover the result $\nu=2$ for the normal distribution.

For $v=\delta_w<0$ (ferromagnetic case), i.e. when the allowed region of the particle increases in time we consider the large $|v|$ limit
and write Eq.(\ref{pnv}) with Eq.(\ref{largeu}) as $P(n,v) \simeq 1 - B(\alpha) \tilde{c}^{-\alpha}+O(\tilde{c}^
{-3\alpha})$ with $B(\alpha)=\Gamma(1+\alpha) \sin(\pi \alpha/2)/\pi \alpha$. Then, in the large $|v|$ limit
$$P(z,v)=-\ln(1-z)-B(\alpha) |v|^{-\alpha} \sum_{n\ge 1} z^n n^{-\alpha}\; ,$$ where the second term is convergent
even at $z=1$. As a consequence the surviving probability remains finite as $n \to \infty$ and we have the
result, $F(n,v)\simeq 1 - {\rm const} |v|^{-\alpha}$, for $|v| \gg 1$. For a small velocity, $0<|v|\ll 1$,
we can estimate $F(n,v)$ by the following reasoning. After $n=n_c$ steps the distance of the adsorbing
site from the starting point, $y_s=v n_c$, will exceed the size of transverse fluctuations of the walk in
Eq(\ref{transv}), $l_{tr} \sim n_c^{1/\alpha}$, with $n_c \sim |v|^{-\nu(\alpha)}$. Then the walker which
has survived until $n_c$-steps with a probability of $n_c^{-1/2}$, will survive in the following
steps with probability $O(1)$. Consequently
\begin{equation}
\lim_{t \to \infty} P_{surv}(n,\delta_w) \sim \lim_{n \to \infty} \left. F(n,v) \right|_{v=\delta_w, n=t}
\sim |\delta_w|^{\nu(\alpha)/2}\;.
\label{psurv-}
\end{equation}
So the average surface magnetization is finite in the
ferromagnetic phase and for small $|\delta|$ it behaves as:
\begin{equation}
\lim_{L \to \infty} \left[m_s(\delta,L)\right]_{\rm av} \sim |\delta|^{\beta_s},\quad \beta_s=\frac{\alpha}{2(\alpha -1)}\;.
\end{equation}
Thus the scaling relation, $\beta_s=x_m^s \nu$, is satisfied.

To end with this section, we will put some words on the typical behaviour and distribution function of the surface magnetization. 
At the critical point, the typical samples, which are represented by absorbed random walks, have a vanishing surface magnetization in the thermodynamic limit. For a large but finite system of size $L$, the surface magnetization is dominated by the largest negative axis excursion of the walker (in the absorbing region), that is by the fluctuation of the walk which scales as $L^{1/\alpha}$, so that
\begin{equation}
m_s^{typ} (L)\sim \exp(-{\rm const.} L^{1/\alpha})
\end{equation}
and the appropriate scaling variable is $\ln m_s/L^{1/\alpha}$. 
In the normal distribution limit $\alpha=2$, we recover the $L^{1/2}$
dependence. 
In ref. \cite{kalin} we have also studied numerically the bulk
magnetic critical behaviour. We have shown that the critical behaviour is controlled by a line of fixed points, where the critical exponents vary continuously with the L\'evy index $\alpha$ up to the normal limit $\alpha=2$, where they take their Fisher's value.

\section{Inhomogeneous disorder}
\subsection{Definition}
Up to now we have only considered homogeneous disorder, that is site independent distributions. But in many physical situations the disorder is not homogeneous. The inhomogeneity can be generated by the presence of a boundary (a surface) or an internal defect which may locally induce a perturbation on the distribution of the couplings and/or fields. 
Let us consider surface induced inhomogeneities which are characterized by a
power-law variation in the probability distribution: $\pi_l(J)-\pi(J)
\sim l^{-\kappa}$ and/or $\rho_l(h)-\rho(h) \sim l^{-\kappa}$, for $l \gg
1$, such that the local control parameter $\delta(l)$ varies as \cite{kaju}
\begin{equation}
\delta(l)=\delta-Al^{-\kappa}\,.
\label{e1.8}
\end{equation}
$\pi(J)$ and $\rho(h)$ are the asymptotic, far from the surface, distributions.
This choice for the functional form of the inhomogeneous disorder is
motivated by the analogy it has with the so-called
extended surface defect problem, introduced and studied by Hilhorst,
van Leeuwen and others, in the two-dimensional classical Ising
model \cite{hilhorst81,blhi83,bugu84,blhi85,buig90}.  This type of inhomogeneity has been later studied for
other models and different geometries. For a review see 
Ref.\cite{igpetu}. Such perturbation is expected to alter only the surface critical behaviour. Due to 
the asymptotic decay, the bulk remain unperturbed by the presence of the inhomogeneity. So that here, the bulk 
critical behaviour is in the Fisher universality class.
On the contrary, depending on the decay exponent $\kappa$, one would expect a modification of the surface critical behaviour. We will illustrate that by considering the surface magnetic behaviour since, thanks to the random walk mapping already discussed, we are able to obtain analytical results \cite{kaju}. 

\subsection{Surface magnetic behaviour}

For the inhomogeneous disorder, the local quantum control parameter has a smooth
position dependence which, at the bulk critical point, is given by
$\delta(l)=-Al^{-\kappa}$ according to Eq.~(\ref{e1.8}). The corresponding
random walk has a locally varying bias with the same type of asymptotic dependence,
$\delta_w(l)=-A_wl^{-\kappa}$. Consequently the average motion of the
walker is parabolic:
\begin{equation}
y_p(t)=-\sum_{l=1}^t\delta_w(l)=\frac{A_w}{1-\kappa}t^{1-\kappa}\,.
\label{e3.3}
\end{equation}
Under the change of variable $y(t)\to y(t)-y_p(t)$, the surviving
probability of the inhomogeneously drifted walker is also the
surviving probability of an unbiased walker, however with a time-dependent
absorbing boundary condition at $y(t)<-y_p(t)$.

The surviving probability of a random walker with
time-dependent absorbing boundaries has already been studied in the
mathematical \cite{uchiyama80} and physical \cite{turban92,kaprivsky96} literature. In a continuum 
approximation, it
follows from the solution of the diffusion equation with appropriate
boundary conditions,
\begin{equation}
\frac{\partial}{\partial t}
P(y,t)=D\frac{\partial^2}{\partial y^2}P(y,t)\,,
\ \ P[-y_p(t),t]=0\,.
\label{e3.4}
\end{equation}
Here $P(y,t)$ is the probability density for the position of the walker at
time $t$ so that the surviving probability is given by
\begin{equation}
P_{\rm surv}(t)=\int_{-y_p(t)}^{\infty}\!\! P(y,t)dy\,.
\label{e3.5}
\end{equation}
The behaviour of the surviving probability depends on the value of the
decay exponent $\kappa$. For $\kappa>1/2$, the drift of the absorbing boundary
in Eq.~(\ref{e3.3}) is slower then the diffusive motion of the
walker, typically given by
\begin{equation}
y_d(t)\sim (D t)^{1/2}\,,
\label{e3.6}
\end{equation}
thus the surviving probability behaves as in the static case. When
$\kappa<1/2$, the drift of the absorbing boundary is faster than the
diffusive motion of the walker and leads to a new behaviour for the surviving
probability. For $A_w>0$, since the distance to the moving boundary grows in
time, the surviving probability approaches a finite limit. On the contrary,
for $A_w<0$, the boundary moves towards the walker and the surviving
probability decreases with a fast, stretched-exponential
dependence on $t$. Finally, in the borderline case $\kappa=1/2$
where the drift of the boundary and the diffusive motion have the same
dependence on $t$, like in the static case the surviving probability decays as
a power, $P_{\rm surv}(\delta_w,t)\sim t^{-\theta(A_w)}$, however with a
continuously varying critical exponent.

\subsection{Critical behaviour with relevant inhomogeneity}

Let us first consider the probability distribution of $\ln m_s$ on finite
samples with length $L$ at the bulk critical point $\delta=0$. According to the arguments given above for the typical magnetization, $[\ln m_s]_{\rm av}$ is expected to scale as
$[\ln\prod_{l=1}^L(J_l/h_l)]_{\rm av}$ when the perturbation tends to reduce
the surface order, i.e., when $A<0$ in Eq.~(\ref{e1.8}). Thus one obtains:
\begin{eqnarray}
[\ln m_s]_{\rm av}&\sim&\sum_{l=1}^L([\ln J_l]
_{\rm av}-[\ln h_l]_{\rm av}\nonumber\\
&=&A\sum_{l=1}^Ll^{-\kappa}\sim AL^{1-\kappa}\,.
\label{e4.1}
\end{eqnarray}
The typical magnetization, defined through
$[\ln m_s]_{\rm av}=\ln[m_s]_{\rm typ}$, has a stretched-exponential
behaviour, 
\begin{equation}
[m_s]_{\rm typ}\sim\exp( {\rm const}\, AL^{1-\kappa})\; ,  A<0\; .
\label{e4.2a}
\end{equation}
For the probability distribution of $\ln m_s$, Eq.~(\ref{e4.1}) suggests the
following scaling form:
\begin{equation}
P_m(\ln m_s, L)=\frac{1}{ L^{1-\kappa}}
\tilde{P}_m\left[{\ln m_s}{L^{1-\kappa}} \right]\,.
\label{e4.2}
\end{equation}
which was successfully tested numerically \cite{kaju}.

The asymptotic behaviour of the scaling function $\tilde{P}_m(v)$ as $v\to0$
depends on the sign of $A$. For enhanced surface couplings, $A>0$, there
is a nonvanishing surface magnetization at the bulk critical point as
$L\to\infty$, thus the powers of $L$ in Eq.~(\ref{e4.2}) must cancel
and $\lim_{v\to0}\tilde{P}_m(v)\sim v^{-1}$. For reduced surface couplings,
$A>0$, $\lim_{v\to0}\tilde{P}_m(v)=0$, indicating a vanishing surface magnetization.

Next we calculate the average behaviour of the surface magnetization, which is determined by the rare events with $m_s=O(1)$. In the random walk picture, starting with $A\sim A_w>0$,
for $A_w\ll1$ one defines the time scale $t^*$ for which the parabolic and
diffusive lengths in Eqs.~(\ref{e3.3}) and~(\ref{e3.6}) are of the same order,
$y_p(t^*)\sim y_d(t^*)$, such that $t^*\sim A_w^{-2/(1-2\kappa)}$. The
surviving probability can be estimated by noticing that, if the walker is not
absorbed up to $t^*$, it will later survive with a finite probability. Thus
$P_{\rm surv}(A_w>0)\sim(t^*)^{-1/2}\sim A_w^{1/(1-2\kappa)}$ and the average
surface magnetization has the same behaviour at the bulk critical point:
\begin{equation}
[m_s]_{\rm av}\sim A^{1/(1-2\kappa)},\ \ \kappa<\frac{1}{2},\ \ 0<A\ll1.
\label{e4.3}
\end{equation}

In the case of a shrinking interval between the walker and the absorbing
wall, $A_w<0$, the leading behaviour of the surviving probability can be estimated by looking for the fraction of walks with
$y(t)>-y_p(t)$ which is given by $P_{\rm surv}(t)\sim\exp[-{\rm const}\,
A_w^2\, t^{1-2\kappa}]$. Thus, for reduced surface couplings, the average
surface magnetization has the following finite-size behaviour at the critical
point:
\begin{equation}
[m_s]_{\rm{av}}\sim\exp\left(-{\rm const}\,A^2\,
L^{1-2\kappa}\right),\kappa<\frac{1}{2},A<0\,.
\label{e4.4}
\end{equation}
One may notice the different size dependence of the typical and
average surface magnetizations at criticality in Eqs.~(\ref{e4.2a})
and~(\ref{e4.4}), respectively.

To obtain the $\delta$-dependence of the average surface magnetization in
the thermodynamic limit, we first determine the typical size of the surface
region $l_s$ which is affected by the inhomogeneity for $\delta<0$ and $A<0$.
It is given by the condition that quantum fluctuation ($\sim\delta$) and
inhomogeneity ($Al^{-\kappa}$) contributions to the energy are of the same
order, from which the relation $l_s\sim\vert A\vert^{1/\kappa}\vert
\delta\vert^{-1/\kappa}$ follows.
Inserting $L\sim l_s$ in Eq.~(\ref{e4.4}), we obtain
\begin{equation}
[m_s]_{\rm{av}}\sim\exp[-{\rm const}\,\vert A\vert^{1/\kappa}
\,\vert\delta\vert^{-(1-2\kappa)/\kappa}],\kappa<\frac{1}{2},A<0\; .
\label{e4.5}
\end{equation}
In ref.~\cite{kaju}, we have also investigated the behaviour of the first gap of the quantum chain and of the surface autocorrelation function. From the first gap study, the relation between time and length scales is found to be
\begin{equation}
\ln \tau \sim \xi^{1-\kappa} \; .
\label{e4.6}
\end{equation}
In order to obtain an estimate for the average surface autocorrelation
function one may notice that the disorder being strictly correlated along the
time axis, a sample with a finite surface magnetization $m_s=O(1)$ has
also a nonvanishing surface autocorrelation function $G_s(\tau)\sim m_s^2$.
Since the fraction of rare events are the same for the two quantities, the
scaling behaviour of $[G_s(L,\tau)]_{\rm{av}}$ can be deduced from the
corresponding relations for the average surface magnetization. For
enhanced surface couplings, $A>0$, according to Eq.~(\ref{e4.3}), $\lim_{\tau
\to\infty}[G_s(\tau)]_{\rm{av}}\sim[m_s(A)]_{\rm{av}}\sim
A^{1/(1-2\kappa)}$ at criticality. For reduced surface couplings, $A<0$, the
finite-size critical behaviour follows from Eq.~(\ref{e4.4}) with
$\lim_{\tau\to\infty}[G_s(L,\tau)]_{\rm av}\sim[m_s(L,A)]_{\rm av}\sim
\exp[-{\rm const}\, A^2\, L^{1-2\kappa}]$. Now, using the scaling
relation (\ref{e4.6}), we obtain
\begin{equation}
[G_s(\tau)]_{\rm av}\sim\exp\left[-{\rm const}\,
A^2\,(\ln\tau)^{\frac{1-2\kappa}{1-\kappa}}\right]\,,
\label{e4.8}
\end{equation}
in the thermodynamic limit.

\subsection{Critical behaviour with marginal inhomogeneity}
In the marginal case $\kappa=1/2$, the differential equation with absorbing boundary conditions in
Eq.~(\ref{e3.4}) can be solved \cite{kaprivsky96} in terms of parabolic cylinder functions of order $\nu$, 
$D_{\nu}(x)$ \cite{abramowitz65}. For zero bulk bias,
$\delta_w=0$, the surviving probability has the asymptotic dependence
$P_{\rm surv}(t)\sim t^{-\vartheta}$ and the exponent $\vartheta$ is such that
\begin{equation}
D_{2\vartheta}(-2A_w)=0\,,
\label{e4.9}
\end{equation}
taking $D=1/2$ for the diffusion constant in Eq.~(\ref{e3.4}). In the limiting cases the solution takes the form
\begin{equation}
\theta=\frac{1}{2}-\sqrt{\frac{2}{\pi}}\, A_w
\label{e4.10}
\end{equation}
for $|A_w|\ll1$ while for large values of $A_w$ one obtains asymptotically:
\begin{equation}
\theta=\sqrt{\frac{2}{\pi}}\,A_w \exp[-2A_w^2]\,.
\label{e4.11}
\end{equation}
Then the random walk correspondence leads to the scaling dimension of the average surface magnetization,
\begin{equation}
x_m^s(A)=\frac{\beta_s}{\nu}=\theta(A_w)\,,
\label{e4.12}
\end{equation}
where one has to identify the relation between the quantum chain perturbation amplitude $A$ and the corresponding random walk drift amplitude $A_w$.  For the extreme binary distribution, one has $A=A_w$, but for other distributions the relation could be different. One has to take a reference point from which the identification is made or one may try to give a convincing enough argument \cite{kaju}. Nevertheless,
the surface magnetization exponent thus depends on the
distribution, however only through the value of the parameter $A_w$ entering
in Eq.~(\ref{e4.12}), the functional form remaining the same.

Next we discuss the properties of the probability distribution of $\ln m_s$.
Repeating the argument used above for relevant perturbations, we arrive to the
scaling form of Eq.~(\ref{e4.2}), however with $\kappa=1/2$. We note that
the same scaling form remains valid for irrelevant perturbations, with
$\kappa>1/2$, but the scaling function $\tilde{P}_m(v)$ has different
limiting behaviours when $v\to0$ in the two cases. While for
$\kappa>1/2$ it approaches a constant, in the marginal case it goes to zero as
$\lim_{v\to0}\tilde{P}_m(v)\sim  v^{-2x_m^s(A)+1}$. In this way one
obtains the proper scaling behaviour for the average surface magnetization. 

For the distribution of the energy gap at the critical point, the scaling relation in Eq.~(\ref{e4.6}) remains valid, with $\kappa=1/2$.

The asymptotic behaviour of the average surface autocorrelation function can
be  determined at criticality by scaling considerations like in a previous
work for the homogeneous case \cite{rieger97b}. Here we make use of the
fact that, as already explained for relevant perturbations, the average
autocorrelation function has the same scaling properties as the average
surface magnetization. Under a scaling transformation when lengths are
rescaled by a factor $b>1$, such that $l'=l/b$, the average surface
autocorrelation behaves as
\begin{equation}
[G_s(\ln\tau)]_{\rm{av}}=b^{-x_m^s}[G_s(\ln\tau/b^{1/2})]_{\rm{av}}\,,
\label{e4.14}
\end{equation}
where we used Eq.~(\ref{e4.6}) to relate the time and
length scales. Taking now $b=(\ln\tau)^2$ we obtain:
\begin{equation}
[G_s(\tau)]_{\rm{av}}\sim(\ln\tau)^{-2x_m^s}\sim(\ln\tau)^{-\beta_s(A)}\,.
\label{e4.15}
\end{equation}
The last expression follows from the exponent relation in Eq.~(\ref{e4.12})
where the average correlation length exponent is $\nu=2$. This scaling
behaviour has also been verified by numerical calculations \cite{kaju}.

\chapter{Non-equilibrium behaviour}

\section{Equations of motion}
In this chapter, we turn to the out of equilibrium dynamical behaviour of the free fermionic chains.  The presentation is of review type and most of the results exposed here were obtain by others. Nevertheless, in order to be as clear as possible, we have presented the topic in a rational way, using the available knowlodge at the moment. Of course, the original references are cited within the text.
Non-equilibrium properties of quantum systems have attracted a lot of interest through
decades \cite{kubo}. Precursor studies on free-fermionic spin chains
were performed by Niemeijer \cite{niemeijer} and Tjion \cite{tjion} at the end of the sixties.
Soon after, Barouch, McCoy and Dresden \cite{bamcdr,bamc} solved exactly the Liouville equation for
the $XY$-chain and computed the time dependent transverse magnetization for several 
non-equilibrium situations.
They have shown the occurrence of an algebraic relaxation instead of exponential as predicted
by Terwiel and Mazur \cite{terwielmazur} using a weak coupling limit.
More recently, a special focus was pointed on non-equilibrium quantum steady states, which 
are driven by some currents \cite{antal1,antal2,eisler}. These studies
are motivated by the fact that quantum systems have a natural dynamics, given by the 
quantum Liouville equation \cite{blum}
$$
\frac{\partial \rho(t)}{\partial t}=-{\rm i} [H_T,\rho(t)]\equiv {\cal L}(\rho(t))\; ,
$$
where $\rho$ is the density operator, $H_T$ the Hamiltonian and where
$\cal L$ is the quantum Liouville super-operator acting on the vector space of linear 
operators. The expectation value of an operator $Q$ at time $t$ is given by
$$
\langle Q\rangle (t)=Tr\{\rho(t)Q\}
$$
where we have assumed that the density operator is normalised to one.
The traditional way to study non-equilibrium properties of a quantum system is to couple 
it with a bath that can be described itself quantum mechanically, for example an assembly 
of harmonic oscillators \cite{caldeira}.
The total Hamiltonian is then splitted into three different pieces:
$$
H_T=H_{s}+H_{I}+H_{b}
$$
where $H_{s}$ is the system Hamiltonian, $H_{b}$ the Hamiltonian of the bath and $H_{I}$ 
stands for the interaction between the bath and system dynamical variables.
Now, average system's quantities are obtained with the help of the reduced density operator, 
that is, density operator traced over the bath dynamical degrees of freedom:
$$
\rho_{s}(t)=Tr_{b}\{\rho(t)\}\; .
$$
The expectation value of a dynamical variable $Q$ of the system is given by
$$
\langle Q \rangle (t)=Tr_{s}\{Q\rho_{s}(t)\}\; .
$$
For a complete review of this approach one can refer to Ref.~\cite{weiss}.

Another, more simple, route to out-of-equilibrium problems, is to investigate the relaxation
of a non-equilibrium initial state, in which a closed system has been prepared at time $t=0$. 
For a pure initial state, $|\Psi\rangle$, the initial density operator is just the projector
$$
\rho=|\Psi\rangle\langle\Psi|\; .
$$
The time evolution of the state $|\Psi\rangle$ is thus simply given by the Schr\"odinger 
equation and is formally given by
$$
|\Psi(t)\rangle =\exp(-{\rm i}Ht)|\Psi\rangle
$$
where $H$ is the Hamiltonian of the closed system.

In the framework of free fermionic models, the easiest way to handle the problem is to 
solve the Heisenberg equations of motion \cite{jacoby1,jacoby2}:
\begin{equation}
\frac{\rm d}{{\rm d}t}X={\rm i}[H,X]\; .
\end{equation}
Since all the spin variables of the quantum chains can be expressed in terms of 
the \index{Clifford operators} Clifford generators $\{\Gamma\}$,
the first step is to solve the equation of motion for these operators. In order to avoid 
unnecessary minus signs, let us redefine the generators in the form of Ref.~\cite{jacoby1}: 
\index{Jordan-Wigner transformation}
\begin{eqnarray}
P_{2n}=X_{n}=\left(\prod_{i=-\infty}^{n-1}\sigma_{i}^{z}\right)
\sigma_{n}^{x}\quad \Big(=(-1)^{n-1}\Gamma_{n}^{1}\Big)\nonumber\\
P_{2n+1}=Y_{n}=\left(\prod_{i=-\infty}^{n-1}\sigma_{i}^{z}\right)\sigma_{n}^{y}\quad
\Big(=(-1)^{n}\Gamma_{n}^{2}\Big)
\end{eqnarray}
where the chain sites run from $i=-N$ to $i=N$ with the thermodynamic limit 
$N\rightarrow \infty$ implicitly taken.
With the anisotropic Hamiltonian \index{$XY$ Hamiltonian}
$$
H=-\frac{1}{2}\sum_{i}\left[J^{x}\sigma_{i}^{x}\sigma_{i+1}^{x}+
J^{y}\sigma_{i}^{y}\sigma_{i+1}^{y}+h\sigma_{i}^{z}\right]
$$
the equations of motion are
\begin{eqnarray}
\frac{\rm d}{{\rm d}t}X_{n}=J^{x}Y_{n-1}+J^{y}Y_{n+1}-hY_{n}\nonumber\\
\frac{\rm d}{{\rm d}t}Y_{n}=-J^{x}X_{n+1}-J^{y}X_{n-1}+hX_{n}\; .
\end{eqnarray}
The solutions are linear combinations of the initial time operators $\{P_{n}\}$:
\begin{eqnarray}
X_{n}=\sum_{m}X_{n-m}f_{m}(t)+Y_{n-m}h_{m}(t)\nonumber\\
Y_{n}=\sum_{m}-X_{n-m}h_{-m}(t)+Y_{n-m}f_{m}(t)
\label{eqfh}
\end{eqnarray}
with
\begin{equation}
f_{m}(t)=\frac{1}{2\pi}\int_{-\pi}^{\pi}{\rm d}q\; \exp(-{\rm i}mq) \cos(\epsilon_{q}t)
\end{equation}
and
\begin{equation}
h_{m}(t)=\frac{1}{2\pi}\int_{-\pi}^{\pi}{\rm d}q\; \exp(-{\rm i}mq) 
\frac{\sin(\epsilon_{q}t)}{\epsilon_{q}}
\left[J^{x}{\rm e}^{{\rm i}q}-h+J^{y}{\rm e}^{-{\rm i}q}\right]
\end{equation}
where the excitation energies are
\begin{equation}
\epsilon_{q}=\left[\left((J^{x}+J^{y})\cos q-h\right)^{2}+
(J^{x}-J^{y})^{2}\sin^{2}q\right]^{1/2}\; .
\end{equation}

In the special cases of the critical Ising chain,  \index{Ising quantum chain} corresponding
to $J^{x}=h=1$ and $J^{y}=0$, and the isotropic $XX$-chain \index{$XX$ quantum chain}  
with $J_x=J_y=1/2$, one obtains closed analytical expressions for the basic time-dependent 
operators \cite{jacoby2}. These closed forms permit to analyse exactly the long time behaviour
of transverse magnetization profiles, \index{Transverse magnetization}
end-to-end correlations \cite{igloirieger,abrietkarevski},
two-time transverse correlation functions and so on. In order to illustrate the procedure 
we concentrate on the critical Ising chain following Ref.~\cite{jacoby1}. The isotropic 
chain can be treated on the same footings. At $J^{x}=h=1$ and $J^{y}=0$, the equations
of motion simplify into
\begin{equation}
\frac{\rm d}{{\rm d}t}P_{n}=P_{n-1}-P_{n+1}\; ,
\end{equation}
where we have used the operator notations $P$ in order to have more compact expressions.
By iteration, one obtains for the $l$th order term
\begin{equation}
\left(\frac{\rm d}{{\rm d}t}\right)^{l}P_{n}=\sum_{k=0}^{l}(-1)^{l}
\left(
\begin{array}{c}
k\\
l
\end{array}
\right)
P_{n-l+2k}\; ,
\end{equation}
where the 
$
\left(
\begin{array}{c}
k\\
l
\end{array}
\right)
$ 
are the binomial coefficients. One can now sum up the Taylor series
\begin{equation}
\sum_{l=0}^{\infty}\frac{z^{l}}{l!}\left(\frac{\rm d}{{\rm d}t}\right)^{l}P_{n}
=\sum_{k=-\infty}^{\infty}P_{n-k} J_{k}(2z)\; ,
\end{equation}
where $J_{n}$ is the Bessel function of integer order $n$. Finally one has
\begin{equation}
P_{n}(t)=\sum_{k=-\infty}^{\infty}P_{k} J_{n-k}(2t)\; ,
\label{eqP}
\end{equation}
which gives the time dependence of the basic operators.
Putting $F_{t}(j)=J_{j}(2t)$, the time evolution of the operators $P_{n}$ are expressed as 
a discrete convolution product:
\begin{equation}
P_{n}(t)=\sum_{k}F_{t}(n-k)P_{k}=(F_{t}\star P)(n)\; ,
\end{equation}
where the kernel for the critical Ising chain  \index{Ising quantum chain} is simply the 
integer Bessel function $J_{k}(2t)$.
This formula is exact for an infinite chain. In the case of a finite
size system, one has to take care for site indices close to the boundaries. We will come 
back later on this point. Alternatively, one can express in closed form the functions $f$ 
and $h$ introduced previously. One has for the critical Ising chain
\begin{eqnarray}
f_{m}(t)=\frac{1}{2\pi}\int_{-\pi}^{\pi}{\rm d}q {\rm e}^{-{\rm i}mq}
\cos\left(2t\sin\frac{q}{2}\right)=J_{2m}(2t)\nonumber\\
h_{m}(t)=\frac{1}{2\pi}\int_{-\pi}^{\pi}{\rm d}q {\rm e}^{-{\rm i}(m-1/2)q}
\sin\left(2t\sin\frac{q}{2}\right)=J_{2m-1}(2t)
\end{eqnarray}
leading together with (\ref{eqfh}) to the time-behaviour (\ref{eqP}).

Having solved explicitly the equations of motion for the Clifford operators, we can now 
express the time dependence of physical quantities in terms of the initial-time $P_{n}$ 
operators. \index{Clifford operators}
For example, the transverse magnetization at site $n$ is given by 
\index{Transverse magnetization}
\begin{equation}
\sigma_{n}^{z}(t)={\rm i}P_{2n+1}(t)P_{2n}(t){\rm i}\sum_{kk'}F_{t}(2n+1-k)F_{t}(2n-k')P_{k}P_{k'}
\end{equation}
which can be rewritten
\begin{eqnarray}
\sigma_{n}^{z}(t)&=&\sum_{p}
\left[F_{t}^{2}(2(n-p))-F_{t}(2(n-p)+1)F_{t}(2(n-p)-1)\right]\sigma_{p}^{z}\nonumber\\
&+&{\rm i}{\sum_{kk'}}^{*}F_{t}(2n+1-k)F_{t}(2n-k')P_{k}P_{k'}
\label{eqsz}
\end{eqnarray}
where ${\sum}^{*}$ restricts the summation over all $k$ and $k'$ except $k'=k\pm 1$. 
The case $k=k'$, with $P_{k}^2=1$,
gives zero since
$$
\sum_{k}F_{t}(2n+1-k)F_{t}(2n-k)=\sum_{k}J_{k}(2t)J_{k+1}(2t)=J_{1}(0)=0\; .
$$
The equation (\ref{eqsz}) will be our starting point for the study of the transverse 
profile time-evolution with specific initial state.

In the same way, the operator $\sigma_{n}^{x}$ at time $t$ is given by
\begin{equation}
\sigma_{n}^{x}(t)=(-{\rm i})^{n}\prod_{j=0}^{2n}\sum_{k_{j}}P_{k_{j}}J_{j-k_{j}}(2t)
\end{equation}
where we have labelled the first site of the chain by $l=0$ and taken a bulk site $n$, 
that is infinitely far away from the boundaries. In the string $\prod_{k_j}P_{k_j}$, 
only those terms with all $k_{j}$ different will give a non vanishing contribution, 
since if at least two identical labels appear, we have a factor
$$
\sum_{k}J_{j-k}(2t)J_{i-k}(2t)=\delta_{ij}
$$
vanishing for $i\neq j$. Finally one arrives at
\begin{equation}
\sigma_{n}^{x}(t)=(-{\rm i})^{n}
\sum_{k_{0}<k_{1}<...<k_{2n}} P_{k_{0}}P_{k_{1}}...P_{k_{2n}} \det(a_{ij})\; ,
\end{equation}
with
\begin{equation}
a_{ij}=J_{i-k_{j}}(2t)\qquad 0\le i,j\le 2n \; .
\end{equation}

The developments given so far are of course valid for infinite homogeneous chains.
In the case of inhomogeneous chains, as for example
in the presence of disorder \cite{abrietkarevski}, the time dependent coefficients of the 
linear development of the operators $P_n(t)$ are no longer given in terms of Bessel 
functions. Moreover, for finite homogeneous chains, they are also terms
proportional not only to $J_{n-k}(2t)$ but also to $J_{n+k}(2t)$, where the last terms 
play a significant role for the near-boundary behaviour.
In order to take into account these facts, we present now quickly the general 
expressions \cite{karevski2} valid for finite chains of the type (\ref{eq1}). 
By introducing the Clifford $\Gamma_{n}^{i}$ operators we arrived at
$H=(1/4)\Gamma^{\dagger}{\bf T}\Gamma$ with the $2L$-component \index{Clifford operators} 
Clifford operator given by 
$\Gamma^{\dagger}=(\Gamma_{1}^{\dagger},\Gamma_{2}^{\dagger},...,\Gamma_{L}^{\dagger})$
and $\Gamma_{n}^{\dagger}=(\Gamma_{n}^{1},\Gamma_{n}^{2})$. The diagonalization of the 
Hamiltonian can then be performed by the introduction of the diagonal Clifford generators
$\gamma_{q}^{\dagger}=(\gamma_{q}^{1},\gamma_{q}^{2})$ related to the lattice one by
$\Gamma_{n}^{1}=\sum_{q}\phi_{q}(n)\gamma_{q}^{1}$ and
$\Gamma_{n}^{2}=\sum_{q}\psi_{q}(n)\gamma_{q}^{2}$ with $\phi$ and $\psi$ defined previously 
as the eigenvector components of the matrix $\bf T$ \index{$\bf T$ Matrix} given in 
(\ref{eqT}). It leads to
\begin{equation}
H={\rm i}\sum_{q} \frac{\epsilon_{q}}{2}\gamma_{q}^{1}\gamma_{q}^{2}\; .
\end{equation}
The time dependence of the diagonal operators is then simply given by
$\gamma_{q}(t)=U_{q}^{\dagger}(t)\gamma_{q}U_{q}(t)$ with
\begin{equation}
U_{q}(t)=\exp\left(\frac{\epsilon_{q}t}{2}\gamma_{q}^{1}\gamma_{q}^{2}\right)=\cos \frac{\epsilon_{q}t}{2}
+\gamma_{q}^{1}\gamma_{q}^{2}\sin\frac{\epsilon_{q}t}{2}\; .
\end{equation}
Utilising the fact that $\{\gamma_{q}^{i},\gamma_{q'}^{j}\}=2\delta_{ij}\delta_{qq'}$, 
we obtain
\begin{equation}
\gamma_{q}^{i}(t)=\sum_{j=1}^{2}\langle \gamma_{q}^{j}|\gamma_{q}^{i}(t)\rangle\gamma_{q}^{j}
\end{equation}
where we have defined the pseudo-scalar product as
\begin{equation}
\langle C|D\rangle= \frac{1}{2}\{C^{\dagger},D\}
\end{equation}
with $\{.,.\}$ the anticommutator. The time-dependent lattice Clifford generators,
$\Gamma_{n}^{i}(t)$, can then be re-expressed in terms of the initial time operators $\Gamma$
with the help of the inverse transforms $\gamma_{q}^{1}=\sum_{k}\phi_{q}(k)\Gamma_{k}^{1}$
and $\gamma_{q}^{2}=\sum_{k}\psi_{q}(k)\Gamma_{k}^{2}$. Finally, one obtains
\begin{equation}
\Gamma_{n}^{j}(t)=\sum_{k,i}\langle \Gamma_{k}^{i}|\Gamma_{n}^{j}(t)\rangle \Gamma_{k}^{i}
\end{equation}
with components \index{Pair contractions}
\begin{eqnarray}
\langle \Gamma_{k}^{1}|\Gamma_{n}^{1}(t)\rangle &=&
 \sum_{q}\phi_{q}(k)\phi_{q}(n)\cos \epsilon_{q}t\nonumber\\
\langle \Gamma_{k}^{1}|\Gamma_{n}^{2}(t)\rangle &=&
\langle \Gamma_{n}^{2}|\Gamma_{k}^{1}(-t)\rangle =-\sum_{q}\phi_{q}(k)\psi_{q}(n)\sin \epsilon_{q}t\nonumber\\
\langle \Gamma_{k}^{2}|\Gamma_{n}^{2}(t)\rangle &=&
 \sum_{q}\psi_{q}(k)\psi_{q}(n)\cos \epsilon_{q}t\; .
\label{eqcomp}
\end{eqnarray}
These general expressions are exact for all finite size free boundaries free fermionic 
quantum chains.

Formally, since $\langle \Gamma_{k}^{i}|\Gamma_{l}^{j}\rangle=\delta_{ij}\delta_{kl}$,
the set $\{\Gamma_{k}^{i}\}$ forms an orthonormal basis of a $2L$-dimensional linear 
vector space $\cal E$ with inner product defined by
$\langle.|.\rangle\equiv \frac{1}{2}\{.^{\dagger},.\}$. Hence, every vector $X\in {\cal E}$ 
has a unique expansion $X=\sum_{i,k}\langle\Gamma_{k}^{i}|X\rangle\Gamma_{k}^{i}$. 
The string expression $X_{1}X_{2}...X_{n}$, with $X_{j}\in {\cal E}$,
is a direct product vector of the space 
${\cal E}_{1}\otimes{\cal E}_{2}\otimes...\otimes{\cal E}_{n}$ which decomposition is
\begin{equation}
X_{1}X_{2}...X_{n}=\sum_{i_{1},k_{1},...,i_{n},k_{n}}
\langle\Gamma_{k_{1}}^{i_{1}}|X_{1}\rangle...
\langle\Gamma_{k_{n}}^{i_{n}}|X_{n}\rangle
\Gamma_{k_{1}}^{i_{1}}...\Gamma_{k_{n}}^{i_{n}}\; .
\label{eqstring}
\end{equation}
With the help of this time-development and the specific solutions $\phi$ and $\psi$, we 
are able to analyse any non-homogeneous finite chain. One has to solve first the eigenvalue 
equation ${\bf T}V_{q}=\epsilon_{q}V_{q}$ and re-inject the $\phi$ and $\psi$ into the basic 
components given in (\ref{eqcomp}) and then use the general expression (\ref{eqstring}).

\section{Time-dependent behaviour}
\subsection{Transverse magnetization}
In this section we consider initial pure states of the form \index{Transverse magnetization}
$$
|\Psi\rangle =|\dots\sigma(k)\sigma(k+1)\dots\rangle
$$
where $\sigma(k)$ is the value of the $z$-component spin at site $k$.
Such initial states are physically relevant since they are quite easily accessible by the 
application of a strong modulated magnetic field in the desired direction. On the 
theoretical side, these states permit to obtain exact solutions \cite{berim1,berim2}.
The question we ask is how the magnetization will relax
as time evolves \cite{berim2,berim,berim3,trimper,igloirieger} and how topological
defects, such as droplets or kinks will spread out \cite{antal1,antal2,trimper,karevski2} ?
In order to give an answer, we compute the transverse
magnetization profile
\begin{equation}
\langle\Psi |\sigma_{l}^{z}(t)|\Psi\rangle = \sum_{k_{1},i_{1},k_{2},i_{2}}
\langle\Gamma_{k_{1}}^{i_{1}}|\Gamma_{l}^{2}(t)\rangle
\langle\Gamma_{k_{2}}^{i_{2}}|\Gamma_{l}^{1}(t)\rangle \langle\Psi|
-{\rm i}\Gamma_{k_{1}}^{i_{1}}\Gamma_{k_{2}}^{i_{2}}|\Psi\rangle\; .
\end{equation}
Since in the $z$-state 
$|\Psi\rangle=\dots\otimes|\sigma(k)\rangle\otimes|\sigma(k+1)\rangle\otimes\dots$
the only non-vanishing contributions come from terms 
$-{\rm i}\Gamma_{k}^{2}\Gamma_{k}^{1}=\sigma_{k}^{z}$, the expectation
value of the transverse magnetization at time $t$ in such a state is given by: 
\index{Transverse magnetization}
\begin{equation}
\langle\Psi |\sigma_{l}^{z}(t)|\Psi\rangle = 
\sum_{k}\left[\langle\Gamma_{k}^{2}|\Gamma_{l}^{2}(t)\rangle
\langle\Gamma_{k}^{1}|\Gamma_{l}^{1}(t)\rangle-
\langle\Gamma_{k}^{1}|\Gamma_{l}^{2}(t)\rangle
\langle\Gamma_{k}^{2}|\Gamma_{l}^{1}(t)\rangle\right]
\sigma(k)\; .
\label{sz1}
\end{equation}
Clearly, for a translation invariant Hamiltonian this equation can be rewritten as a 
discrete convolution product \cite{karevski2}:
\begin{equation}
m(l,t)=\sum_{k}G_{t}(l-k)\sigma(k)=\big(G_{t}\star\sigma\big)(l)\; ,
\label{eqconv}
\end{equation}
with $m(l,t)=\langle\Psi |\sigma_{l}^{z}(t)|\Psi\rangle$.
As it is seen from (\ref{eqsz}) for the critical infinite Ising chain 
\index{Ising quantum chain}  the kernel $G_{t}$ is given in
terms of Bessel functions:
\begin{equation}
G_{t}(l)=F^{2}_{t}(2l)-F_{t}(2l+1)F_{t}(2l-1)
\end{equation}
with $F_{t}(n)=J_{n}(2t)$ as already defined.
Due to the different asymptotic properties of the Bessel functions one has to distinguish 
between the cases $n/t=v>1$ and $v<1$. For $v>1$, corresponding in (\ref{eqconv}) to a 
distance $n=l-k$ between sites $l$ and  $k$ larger than $t$, we are in the acausal region 
since the elementary excitations, travelling with velocity equal to one by appropriate
normalisation of the Hamiltonian (\ref{eq1}), have no time to propagate from the initial 
position $l$ up to site $k$. This is exactly what is seen from the asymptotic behaviour of 
the Bessel function $J_{n}(t)$ which is vanishing exponentially as $\exp(-\lambda(v) n)$ 
with $\lambda(v)>0$ for $n>t$. So the behaviour of the local magnetization will
be completely governed by the local environnement, since we have a compact support kernel, 
insuring the existence of the convolution product. Inside the causal region, $v<1$, 
with the help of the asymptotic \cite{gradstein} for $\nu\gg 1$
\begin{equation}
J_{\nu}\left(\frac{\nu}{\cos\beta}\right)=\sqrt{\frac{2}{\pi \nu\tan \beta}}\cos\delta
\end{equation}
where $\delta\equiv \nu (\tan \beta-\beta)-\pi/4$, one obtains in the continuum limit
\begin{equation}
G_{t}(vt)=\frac{1}{t}g(v)
\end{equation}
with
\begin{equation}
g(v)=\left\{\begin{array}{ll}
\frac{1}{\pi} \left( 1-v^{2} \right)^{1/2}&\quad |v|<1 \\
0&\quad |v|>1
\end{array}
\right.\; .
\label{kernelgi}
\end{equation}
In the continuum limit, the local magnetization $m(n,t)=m_{t}(v)$ is then given by the 
convolution product
\begin{equation}
m_{t}(v)=\big(g\star \sigma_{t}\big)(v)
\end{equation}
with the initial state function $\sigma_{t}(v)=\sigma(tv)$.

The same analysis for the $XX$-chain \cite{karevski2}  \index{$XX$ quantum chain} leads for 
the Green function $g(v)$ to
\begin{equation}
g(v)=\left\{\begin{array}{ll}
\frac{1}{\pi} \left( 1-v^{2} \right)^{-1/2}&\quad |v|<1 \\
0&\quad |v|>1
\end{array}
\right.\; ,
\label{kernelgxx}
\end{equation}
that is an inverse square-root behaviour. From these expressions one is able to evaluate 
the long time behaviour of the relaxation process from any initial $z$-state. With a 
homogeneous initial state, $m(0)=1$, one obtains for the $XX$-chain $m(t)=m(0)=1$ as it 
should be, since the dynamics is conservative. On the contrary, in the Ising case, one has
from (\ref{eqconv}) together with (\ref{kernelgi}) $m(t)=1/2$, that is in the long time 
regime half of the initial magnetization remains in the $z$-direction. A more careful 
analysis\footnote{The continuum analysis performed here is valid up to order $t^{-1}$.}
leads to \cite{trimper,igloirieger}
\begin{equation}
m(t)=\frac{1}{2}+\frac{1}{4t}J_{1}(4t)
\end{equation}
so that the final constant is in fact reached with a power law behaviour $t^{-3/2}$.
One remarkable property of the relaxation of the magnetization of the Ising chain 
\index{Ising quantum chain}  is that the remaining half magnetization has a conservative 
dynamics. For example, if we start with a droplet of $L$ down spins inside an environment
of up spins, in the long time limit $t\gg L$, the magnetization profile is given by
\begin{equation}
m_{t}(v)=g(v)\star \left[1-\frac{2L}{t}\delta(v)\right] = \frac{1}{2}-\frac{2L}{\pi t}\sqrt{1-v^{2}}\;.
\end{equation}
The excess magnetization $m^c=m-1/2$ is a scaling function and spreads out into the bulk 
of the up-spins without loosing any weight. In fact, one can write down a continuity 
equation $\partial_{t}m^c(n,t)+j(n,t)-j(n-1,t)=0$ where the current density is given in 
the continuum limit by \cite{karevski2}
\begin{equation}
j(x,t)=\frac{x}{t}m^c(x,t)\;.
\end{equation}

For a general initial $z$-state the Fourier transform of equation (\ref{eqconv}) is
\begin{equation}
\tilde{m}_t(q)=\tilde{\sigma}_t(q)\tilde{g}(q)
\end{equation}
where the kernel in Fourier space is
\begin{equation}
\tilde{g}(q)=J_0(2\pi q)
\end{equation}
for the $XX$ chain and \index{$XX$ quantum chain} 
\begin{equation}
\tilde{g}(q)=\frac{J_1(2\pi q)}{2\pi q}
\end{equation}
for the Ising one. \index{Ising quantum chain}  By inverse Fourier transform one obtains 
the desired magnetization profile.

It is to be noted here that  in the $XX$ case  an additional step-like structure 
arises on top of the scaling function \cite{hunyadi}. It was found that this step-like structure is related to 
the flow of localized flipped spins. Indeed, each step  width broadens as $t^{1/3}$ while the height decreases
as $t^{-1/3}$, keeping the transported magnetization within each step equal to one unit. 
One may notice that the relaxation of spatially inhomogeneous initial states has been 
treated for several variants of the $XY$ quantum chains, modulated, dimerised and so 
on \cite{cabrera}. A slowing down of the relaxation may occur
for fermionic models that have a gaped excitation spectrum at some special points of
the modulation wave vector \cite{berim,berim3,cabrera}.
Such study was also extended, by time-dependent DMRG method, to the $XXZ$-chains \cite{gobert}. Starting with a kink in the $z$-direction,
it was shown that the magnetization transport remains balistic with a scaling function $\Phi(n/t)$ for small $S^zS^z$ interaction
while it vanishes for strong enough interaction.
Disordered Quantum chains have been investigated numerically in Ref.~\cite{abrietkarevski}.
For the transverse magnetization, they show similar \index{Transverse magnetization}
behaviour to the homogeneous chain, that is an algebraic decay in time.
During the time evolution, the spatial correlations are building up and at long time 
they reach a size-dependent constant depending on the distance from 
criticality \cite{abrietkarevski}. This behaviour is related to the distribution of rare 
samples that are strongly correlated due to large domains of strong couplings.

\subsection{Boundary effects}
For an open chain, one has to take care about boundary effects that could for a 
near-boundary spin modify the relaxation behaviour compared to the bulk one. We will 
illustrate this with the behaviour of the transverse magnetization
in the case of the $XX$ chain.  \index{$XX$ quantum chain} Consider as an initial state 
a droplet of $L$-down spins (in the $z$ direction) at the boundary of a semi-infinite chain, 
the rest pointing in the opposite direction. The transverse magnetization at site
$l$ and at time $t$ can be written \index{Transverse magnetization}
\begin{equation}
m(l,t)=1-2\sum_{k=1}^L F_t (l,k)
\end{equation}
with the Green function $F_t(l,k)$, using the symmetry properties of the basic contractions, 
\index{Pair contractions} given by
\begin{eqnarray}
F_t(l,k)&=&|\langle\Gamma_k^1|\Gamma_l^1(t)\rangle|^2+
|\langle\Gamma_k^1|\Gamma_l^2(t)\rangle|^2\; \nonumber\\
&=&[J_{l-k}(t) - (-1)^{k}J_{l+k}(t)]^2
\end{eqnarray}
which is exactly the form given in~(\ref{zzinfini}) for the $z-z$ correlator
$\langle\sigma_l^z(t)\sigma_ k^z \rangle$ at infinite temperature. The reason for this 
coincidence lies in the fact that the expectations of the string operators $\prod_i P_{2i}$
for the correlator at infinite temperature appear in the very same form in the $z$-state
expectations of the transverse magnetization.
So that, this is also true for the Ising chain. From these coincidence, we see that the 
long time behaviour of the Green function is $F_t\sim t^{-3}$ near the boundary. 
In fact, from the asymptotic analysis of~(\ref{zzinfini}) we can calculate explicitly
the long time behaviour of the magnetization. The droplet magnetization, 
$M^z(t)=(1/2)\sum_{k=1}^{L}\sigma_k^z(t)$ spreads in the bulk, for $t\gg L$, as
\begin{equation}
M^z(t)=\frac{L}{2}-\frac{1}{9\pi}\left(\frac{L^2}{t}\right)^3\; ,
\end{equation}
which as to be compared to \cite{trimper}
\begin{equation}
M^z(t)=\frac{L}{2}-\frac{1}{\pi}\frac{L^2}{t}
\end{equation}
for a bulk droplet.

\subsection{Two-time functions}
Two-time functions \index{Two-time functions}
$\langle {\cal Q}(t_1){\cal Q}(t_2)\rangle$ are of primary importance in characterising 
non-equilibrium dynamics. In particular, they show the phenomenon of aging, \index{Aging} 
that is, the dependence of the correlation functions on both times $t_1$ and $t_2>t_1$, 
where $t_1$ is usually called the waiting time and specify the age of the system. This is 
in contrast to the equilibrium situation where the dependence is only on the time difference 
$t_2-t_1$. Usually, at large waiting times, two distinct regimes develop: {\it (i)} at short 
time differences the correlations are time translation invariant whereas {\it (ii)} at
long time differences, the relaxation is very slow and depends on the waiting time too. 
Aging was first considered in ultra-slow glassy dynamics \cite{bouchaud} but it has been 
also investigated in simpler systems, either classical \cite{cugliandolo2}
as well as quantum \cite{cugliandolo,trimper,pottier,igloirieger}.
The first attempt in this direction on homogeneous short range quantum spin chain 
was done in Ref.~\cite{trimper}.

The two-time non-equilibrium correlation function we consider is defined by
\begin{equation}
C_{ij}(t_{1},t_{2})=\langle \Psi|\sigma_{i}^{z}(t_1)\sigma_{j}^{z}(t_2)|\Psi\rangle\; ,
\end{equation}
with a $z$-initial state $|\Psi\rangle$ as already defined and $t_2>t_1$.
In terms of the time-independent basic operators, the two-spins product 
$\sigma_{i}^{z}(t_1)\sigma_{j}^{z}(t_2)$
takes the form
\begin{equation}
\sigma_{i}^{z}(t_1)\sigma_{j}^{z}(t_2)=-\sum_{i_{1}k_{1}\dots i_{4}k_{4}} 
C_{k_{1},k_2,k_3,k_4}^{i_1,i_2,i_3,i_4}(t_1,t_2)
\Gamma_{k_1}^{i_1}\Gamma_{k_2}^{i_2}\Gamma_{k_3}^{i_3}\Gamma_{k_4}^{i_4}
\end{equation}
where the coefficients of that development are given by applying formula (\ref{eqstring}) 
with
$\sigma_{i}^{z}(t_1)\sigma_{j}^{z}(t_2)=-\Gamma_{i}^{2}(t_1)\Gamma_{i}^{1}(t_1)\Gamma_{j}^{2}(t_2)\Gamma_{j}^{1}(t_2)$.
Using Wick's theorem, the expectation in the $|\Psi\rangle$ state is given by 
\index{Wick Theorem}
\begin{equation}
C_{ij}(t_{1},t_{2})=\sum_{k=0}^{6}T_{k}^{ij}(t_1,t_2)+
\langle\Psi|\sigma_{i}^{z}(t_1)|\Psi\rangle\langle\Psi|\sigma_{j}^{z}(t_2)|\Psi\rangle\; ,
\end{equation}
with a time-translation invariant element
\begin{equation}
T_{0}^{ij}(t_1,t_2)=<A_{i}A_j>^{t_2-t_1}<B_iB_j>^{t_2-t_1}-
<B_iA_j>^{t_2-t_1}<A_iB_j>^{t_2-t_1}
\end{equation}
and non-invariant terms
\begin{eqnarray}
T_{1}^{ij}(t_1,t_2)&=&-<A_{i}A_j>^{t_2-t_1}[BB]^{t_1,t_2}_{i,j}\\
T_{2}^{ij}(t_1,t_2)&=&<A_{i}B_j>^{t_2-t_1}[BA]^{t_1,t_2}_{i,j}\\
T_{3}^{ij}(t_1,t_2)&=&<B_{i}B_j>^{t_2-t_1}[AA]^{t_1,t_2}_{i,j}\\
T_{4}^{ij}(t_1,t_2)&=&-<B_{i}A_j>^{t_2-t_1}[AB]^{t_1,t_2}_{i,j}\\
T_{5}^{ij}(t_1,t_2)&=&-[AA]^{t_1,t_2}_{i,j}[BB]^{t_1,t_2}_{i,j}\\
T_{6}^{ij}(t_1,t_2)&=&[AB]^{t_1,t_2}_{i,j}[BA]^{t_1,t_2}_{i,j}\; ,
\end{eqnarray}
with
\begin{equation}
[CD]^{t_1,t_2}_{i,j}=\sum_{k}\sigma_{k}
\left[<B_kC_i>^{t_1}<A_kD_j>^{t_2}-<A_kC_i>^{t_1}<B_kD_j>^{t_2}\right]
\end{equation}
where we have used the short notations of Ref.~\cite{igloirieger} that is to recall, 
$A_{i}=\Gamma_{i}^{1}$,
$B_{i}=-{\rm i}\Gamma_{i}^{2}$ and for the basic time contractions \index{Pair contractions}
$<XY>^{t}\equiv \langle X|Y(t)\rangle=\frac{1}{2}\{X^{\dagger},Y(t)\}$ that can be read from
Eq.~(\ref{eqcomp}). The elements $T_{k}^{ij}(t_1,t_2)$ satisfy the set of symmetry relations
\begin{equation}
\begin{array}{l}
T_0^{ij}(t_1,t_2)=T_0^{ji}(t_2,t_1)\\
T_1^{ij}(t_1,t_2)=-T_1^{ji}(t_2,t_1)\\
T_2^{ij}(t_1,t_2)=-T_4^{ji}(t_2,t_1)\\
T_3^{ij}(t_1,t_2)=-T_3^{ji}(t_2,t_1)\\
T_5^{ij}(t_1,t_2)=T_5^{ji}(t_2,t_1)\\
T_6^{ij}(t_1,t_2)=T_6^{ji}(t_2,t_1)\; .
\end{array}
\end{equation}
From the previous equations, together with the symmetry relations, one obtains for the 
connected symmetrised spin-spin correlation,
\begin{equation}
\tilde{C}_{ij}(t_1,t_2)=\frac{1}{2}\langle\Psi|\{\sigma_i^z(t_1),\sigma_j^z(t_2)\}|\Psi\rangle-
\langle\Psi|\sigma_{i}^{z}(t_1)|\Psi\rangle\langle\Psi|\sigma_{j}^{z}(t_2)|\Psi\rangle\; ,
\end{equation}
the simpler expression
\begin{equation}
\tilde{C}_{ij}(t_1,t_2)=T_0^{ij}(t_1,t_2)-[AA]^{t_1,t_2}_{i,j}[BB]^{t_1,t_2}_{i,j}+
[AB]^{t_1,t_2}_{i,j}[BA]^{t_1,t_2}_{i,j}\; ,
\end{equation}
which is valid for all free fermionic spin chains, prepared in an initial $z$-state.
One may notice that for an initial $x$-state \cite{igloirieger}, the formula for the 
correlation function is the same, apart that one has to replace the initial definition 
of the $[CD]$ \cite{igloirieger}.

\subsection{Critical Ising chain}
We concentrate now on the critical Ising chain,  \index{Ising quantum chain} since at 
this point the contractions are simpler and one is able to give exact closed analytical 
expressions \cite{igloirieger}.

In particular, the two-time autocorrelation function, that is, \index{Two-time functions}
$i=j$, takes a very simple form for a completely ordered $z$-initial state. Together with
$<A_iA_j>^t=(-1)^{i+j}J_{2(i-j)}(2t)$ and $<A_iB_j>^t={\rm i}(-1)^{i+j+1}J_{2(i-j)+1}(2t)$ 
for the infinite chain, one obtains
\begin{equation}
\tilde{C}(t_1,t_2)=J_0^2(2(t_2-t_1))-\frac{1}{4}[f(t_2+t_1)-g(t_2-t_1)]^2
\end{equation}
with $f(x)=J_2(2x)+J_0(2x)$ and $g(x)=J_2(2x)-J_0(2x)$. The two-time autocorrelation can 
be rewritten
\begin{equation}
\tilde{C}(t_1,t_2)=J_0^2(2(t_2-t_1))-
\left[\frac{J_{1}(2(t_2+t_1))}{2(t_2+t_1)}+J'_1(2(t_2-t_1))\right]^2\; .
\end{equation}
The dependence  of the correlation function on both $t_1$ and $t_2$ reflects the 
non-equilibrium behaviour of the system. However, in the long-time regime $t_1\gg 1$ with 
small difference times $\tau=t_2-t_1\ll t_1$ one recovers time translation invariance, 
that is, the two-time function depends only on the difference $\tau$ and decays
algebraically as $\tau^{-2}$. This power law decay subsists with the same exponent for 
any value of the transverse field $h$ \cite{igloirieger}.

By Kubo's formula \cite{kubo}, the linear response function $R_{ij}^{zz}(t_1,t_2)$ at 
site $j$ to an infinitesimal transverse field at site $i$ is given by the commutator
\begin{equation}
{\rm i}\langle\Psi|[\sigma_{i}^z(t_1),\sigma_j^z(t_2)]|\Psi\rangle
\end{equation}
 and, on the same 
lines as before, it is expressed at the critical field as
\begin{equation}
R_{ij}^{zz}(t_1,t_2)=Q_{ij}^{1}(t_1,t_2)+Q_{ij}^2(t_1,t_2)
\end{equation}
with
\begin{eqnarray}
Q_{ij}^1(t_1,t_2)&=&J_{2r-1}(2\tau)J_{2r+1}(2(t_1+t_2))\frac{2r+1}{t_1+t_2}\nonumber\\
&-&J_{2r+1}(2\tau)J_{2r-1}(2(t_1+t_2))\frac{2r-1}{t_1+t_2}
\end{eqnarray}
and a time translation invariant element
\begin{equation}
Q_{ij}^{2}(t_1,t_2)=Q_{ij}^{2}(\tau)=-\frac{2}{\tau}J_{2r-1}(2\tau)J_{2r+1}(2\tau)
\end{equation}
with $r=j-i$ the distance between the two sites $i$ and $j$.
Translation invariance in time is restored at small difference times, 
$1\ll\tau=t_2-t_1\ll t_1$, since the dominant term is then $Q_{ij}^{2}(\tau)$ and 
it leads to a $\tau^{-2}$ behaviour.

\subsection{XX chain}
For the $XX$-chain,  \index{$XX$ quantum chain} the dynamics is conservative, that is 
the total magnetization in the $z$ direction is conserved during time evolution. 
Therefore, in order to consider a relaxation process one cannot use the same procedure as in
the Ising case. Instead of an infinite chain fully magnetized in the up-direction, we take 
as the initial state of the infinite chain a droplet of $L$-spins pointing upward in 
$z$-direction, from site labels $i=0$ to $i=L-1$, the rest of the spins pointing in the 
opposite direction \cite{trimper}. This can be assimilated to a system of size $L$
interacting through its boundaries with a bath, where it can dissipates magnetization. 
The final stationary state reached by the system is the completely magnetized state, 
with all the system's spins down, which is not the equilibrium state.

The two-time connected autocorrelation function of the droplet is defined by 
\index{Two-time functions}
\begin{equation}
\tilde{C}(t_1,t_2)=\langle \Psi |M^{z}(t_1)M^{z}(t_2)|\Psi\rangle-
\langle \Psi |M^{z}(t_1)|\Psi\rangle\langle \Psi |M^{z}(t_2)|\Psi\rangle
\label{cxx2}
\end{equation}
with $M^{z}(t)=\frac{1}{2}\sum_{i=0}^{L-1}\sigma_{i}^{z}(t)$ the total transverse 
magnetization of the droplet subsystem.
The $1/2$ factor has been introduced in order to follow Ref.~\cite{trimper}.
The Heisenberg equations are easily solved in this case and one obtains for the 
lattice fermions
\begin{equation}
c_{k}(t)=e^{-{\rm i}ht}\sum_{r}{\rm i}^{k-r}J_{k-r}(t)c_{r}\; .
\end{equation}
Introducing these solutions inside (\ref{cxx2}), using Wick theorem and the initial droplet 
state, one obtains
\begin{eqnarray}
\tilde{C}(t_1,t_2)=\sum_{k,l=0}^{L-1}J_{k-l}(t_2-t_1)
\sum_{r=0}^{L-1}J_{k-r}(t_2)J_{l-r}(t_1)\nonumber\\
-\sum_{k,l,r,s=0}^{L-1}J_{k-r}(t_2)J_{k-s}(t_2)J_{l-r}(t_1)J_{l-s}(t_1)\; .
\end{eqnarray}
The long-time behaviour of that expression is obtained with the help of the asymptotic 
expansion of the Bessel function
$J_{n}(t)\sim\sqrt{2/\pi t}\cos(t-n\pi/2-\pi/4)$. For $t_1,t_2\gg L$, one obtains the 
asymptotic form
\begin{equation}
\tilde{C}(t_1,t_2)=\frac{L^3}{2\sqrt{\pi^3 t_1t_2(t_2-t_1)}}-\frac{L^4}{2\pi^2 t_1t_2}\; .
\end{equation}
In the finite magnetization density regime \cite{trimper}, that is for times smaller than 
$L^2$, the second term dominates and one has finally
\begin{equation}
\tilde{C}(t_1,t_2)=-\frac{L^4}{2\pi^2 t_1t_2}t_1^{-2}f\left(\frac{t_2}{t_1}\right)\qquad L\ll t_1,t_2\ll L^2\; ,
\end{equation}
with $f(x)\sim x^{-1}$. One may notice that this scaling form is specific of 
aging \index{Aging}  at a critical point where critical coarsening takes 
place \cite{bray,godreche,henkel2}.
This means that aging occurs in the ordered isotropic $XX$-chain  \index{$XX$ quantum chain} 
with deterministic quantum dynamics \cite{trimper}.
In fact, to conclude positively about the occurrence of aging in such system,
Sch\"utz and Trimper \cite{trimper} have restricted their definition of aging to the 
phenomenon of increasing relaxation times, {\it i. e.} the longer one waits, the slower 
the relaxation becomes.
For that purpose they considered the ratio
\begin{equation}
R(t_1,t_2)=\tilde{C}(t_1,t_2)/\tilde{C}(t_1,t_1)=t_1/t_2\; ,
\end{equation}
that is a kind of normalized autocorrelation function with respect to the waiting 
time $t_1$, which clearly shows that aging occurs in this homogeneous deterministic 
quantum system.

\chapter*{Discussion and summary}
\addcontentsline{toc}{chapter}{Discussion and summary}

We have presented in this review the general ``almost'' canonical approach to free fermionic
quantum spin chains. Almost canonical in the sense that, 
several variants of notations and approaches appear
all along the enormous literature
devoted to this topic,  and one is sometimes confused. However, we tried to give a systematic analysis and emphasise the basic 
technical tools, such as the Wick expansion, pair contractions and operator time development.

We have presented very few results on the well known static properties of these spin chains 
but merely insisted on the way to calculate them, in this sense it is a pedagogical attempt.
Moreover, we have shown explicitly how one can reconstruct
the zero temperature quantum phase diagram from the study of the boundary magnetization. 
Since it is very easy to compute the first site magnetization \cite{peschel,karevski1}, one has paid 
a very low price for the knowledge of the bulk behaviour, compared to standard approaches \cite{barouch}.
Of course, the present analysis was {\it a posteriori} and, as everyone knows, it is an easy task and sometimes pathetic to recover old results by new approaches.
However, one can imagine to use this bias to tackle really new systems, so to say to infer bulk behaviour from surface properties.

Results on equilibrium time-dependent correlation functions are quickly reviewed and I apologise for important works that have not been cited here for various reasons, 
the main being fortuitous omission or lack of knowledge. One should not worry too much 
about it, since the science is, or should be, a public good.
Phrased differently, intellectual property is still a property.
To summarise on the spin-spin time dependent correlations, the calculation of the transverse functions is an easy task, since they have a local expression in terms of 
the non-interacting fermions. They are basically fermion density
correlation functions and show up power law decay in time. The $x$ spin component 
correlations are much more difficult to evaluate,
since in this case the Jordan-Wigner transformation leads to string Fermi operators. 
Nevertheless, using Wick's theorem their asymptotic behaviour can be extracted. At 
vanishing temperature, they decay with a power law whereas at infinite
temperature they have a Gaussian shape. In the intermediate temperature regime, 
the decay is exponential with a power law prefactor.

After the generally known properties of the homogeneous system, we have presented some results on aperiodically modulated systems and random ones. 
In the aperiodic case, we have briefly given a relevance criterion for such perturbation and emphasized the anisotropic scaling behaviour found in such systems as well as the weak universality observed for some of them.
In the random case, in a first part we summirized RG Fisher's results for the homogeneous normally distributed random Ising quantum chain. In the remaining part, we have presented some analytical results obtained in cases where the disorder is no more homogeneous but presents a power law spatial variation or, in a second case, where the disorder is no more normally distributed but rather has long tails of L\'evy type.

Finally, we dealt with some of the non-equilibrium properties 
of these systems. We focused our attention, after solving the Heisenberg equations of motions
for the Fermi operators, on the relaxation of the transverse magnetization in the critical 
Ising chain and the isotropic $XY$ chain. For some particular initial
states, magnetized in a given direction, we derived analytically a general expression that 
can be used to calculate at later times the transverse magnetization profile.  In fact, the 
long time relaxation of the magnetization is expressed as a convolution product of the 
initial state with a response kernel obtained analytically for both  $XX$
and Ising chain. It is seen in particular that the Ising chain exhibits some similarities
with the conserved dynamics $XX$ chain. That is, after a transient regime, the relaxation 
process in the Ising case is conservative too. Two-time functions are also discussed in the 
light of aging  \index{Aging} phenomena. It is shown how, starting with some non-equilibrium 
initial state, the correlations depend explicitly on both the waiting time, characterising
the age of the system, and the later measurement time. Although aging was first considered 
in complex systems such as structural glasses or spin glasses, it is in fact also present 
in simple homogeneous systems with a completely deterministic dynamics.
This final result is in fact very recent and it has been demonstrated in 
classical \cite{cugliandolo2} and quantum systems \cite{trimper}.

\bibliographystyle{unsrt}

\bibliography{biblio}

\end{document}